\pdfoutput = 1
\documentclass[12pt]{scrartcl}
\usepackage[utf8]{inputenc}

\areaset{158mm}{238mm}
\usepackage{setspace}
\usepackage[all]{xypic}
\usepackage{scrlayer-scrpage}

\usepackage{multirow}
\usepackage{amsmath}
\usepackage{mathtools}
\usepackage{mathrsfs}

\usepackage{amssymb}
\usepackage[unicode=true,linktoc=all]{hyperref}
\usepackage{cite}
\usepackage{bm}
\usepackage{slashed}

\usepackage{tikz,pgf}
\usetikzlibrary{shapes}
\usetikzlibrary{calc}
\usetikzlibrary{decorations.pathmorphing}
\usetikzlibrary{decorations.pathreplacing,shapes.misc}
\usetikzlibrary{positioning}
\usetikzlibrary{decorations.markings}
\tikzset{->-/.style={decoration={
  markings,
  mark=at position .5 with {\arrow{>}}},postaction={decorate}}}
\tikzset{-<-/.style={decoration={
  markings,
  mark=at position .5 with {\arrow{<}}},postaction={decorate}}}

\usepackage{xcolor}
  \definecolor{rblue}{RGB}{81, 49, 193}
  \definecolor{rorange}{RGB}{255, 147, 40}
  \definecolor{rgreen}{RGB}{176, 233, 0}

\newcommand{\fsu}{\mathfrak{su}}
\newcommand{\fso}{\mathfrak{so}}
\newcommand{\fsp}{\mathfrak{sp}}

\newcommand{\fe}{\mathfrak{e}}

\newcommand\be{\begin{equation}}
\newcommand\ee{\end{equation}}
\renewcommand{\hat}{\widehat}

\newcommand\SEIT[8]{
    \def\tempa{#1}%
    \def\tempb{#2}%
    \def\tempc{#3}%
    \def\tempd{#4}%
    \def\tempe{#5}%
    \def\tempf{#6}%
    \def\tempg{#7}%
    \SEITc#8
 }
 
 \newcommand\SEITc[3]{
     \def\tempi{#1}%
     \def\tempj{#2}%
     \SEITcc#3
 }    

\newcommand\SEITcc[9]
{$\xymatrix@=0.01pc{&&&\tempi \atop \qquad\\
&&&\overset{\mathfrak{\fsp}_{#8}}{1}\\ &\tempa\,\,\overset{\mathfrak{\fso}_{#1}}{4}&\underset{\tempb}{\overset{\mathfrak{\fsp}_{#2}}{1}}&\underset{\tempc}{\overset{\mathfrak{\fso}_{#3}}{4}}&\underset{\tempd}{\overset{\mathfrak{\fsp}_{#4}}{1}}&\underset{\tempe}{\overset{\mathfrak{\fso}_{#5}}{4}}&\underset{\tempf}{\overset{\mathfrak{\fsp}_{#6}}{1}}&\underset{\tempg}{\overset{\mathfrak{\fso}_{#7}}{4}}&\overset{\mathfrak{\fsp}_{#9}}{1}\tempj}$}

\newcommand\SEsix[6]
{
    \def\tempa{#1}%
    \def\tempb{#2}%
    \def\tempc{#3}%
    \def\tempd{#4}%
    \def\tempe{#5}%
    \SEsixc#6
 }
 
 \newcommand\SEsixc[5]
 {$\begin{matrix}
\tempa\, \overset{\mathfrak{sp}_{#1}}{1} \,\, \underset{\tempb}{\overset{\mathfrak{so}_{#2}}{4}} \,\, \underset{\tempc}{\overset{\mathfrak{sp}_{#3}}{1}} \,\, \underset{\tempd}{\overset{\mathfrak{su}_{#4}}{2}} \,\,\overset{\mathfrak{su}_{#5}}{2} \,\, \tempe
\end{matrix}$}

\newcommand\SEse[9]
{
    \def\tempa{#1}%
    \def\tempb{#2}%
    \def\tempc{#3}%
    \def\tempd{#4}%
    \def\tempe{#5}%
    \def\tempg{#6}%
    \def\temph{#7}%
    \def\tempi{#8}%
    \SEsec#9
 }
 
 \newcommand\SEsec[8]
 {$\begin{matrix}
&&&\tempi\\
&&&\overset{\mathfrak{sp}_{#7}}{1}\\
\tempa \,\,\overset{\mathfrak{sp}_{#8}}{1}&\underset{\tempb}{\overset{\mathfrak{so}_{#1}}{4}}&\underset{\tempc}{\overset{\mathfrak{sp}_{#2}}{1}}&\underset{\tempd}{\overset{\mathfrak{so}_{#3}}{4}}&\underset{\tempe}{\overset{\mathfrak{sp}_{#4}}{1}}&\underset{\tempg}{\overset{\mathfrak{so}_{#5}}{4}}&\overset{\mathfrak{sp}_{#6}}{1} \,\,\temph \\
\end{matrix}$}

\numberwithin{equation}{section}
 
\title{Back to Heterotic Strings \\ on ALE Spaces} 

\subtitle{Part I -- Instantons, 2-groups and T-duality}

\author{Michele Del Zotto$^{\dagger\sharp}$, Muyang Liu $^\sharp$, \\ and Paul-Konstantin Oehlmann$^\dagger$
\\
	\footnotesize\slshape$^\sharp$ Department of Mathematics, Uppsala University, Uppsala, Sweden\\
	\footnotesize\slshape$^\dagger$ Department of Physics and Astronomy, Uppsala University, Uppsala, Sweden\\
}
\date{}

\begin{document}

\maketitle 
\begin{picture}(0,0)
\put(330,290){  UUITP--40/22 }
\end{picture}
\vspace{-13.5cm}
\vspace{14.5cm}

\paragraph{\hspace{.9cm}\large{Abstract}}
\vspace{-.1cm}
\begin{abstract}
\noindent In this paper we begin revisiting the little string theories (LSTs) which govern the dynamics of the instantonic heterotic $E_8 \times E_8$ five-branes probing ALE singularities, building on and extending previous results on the subject by Aspinwall and Morrison as well as Blum and Intriligator.  Our focus are the cases corresponding to choices of non-trivial flat connections at infinity. The latter are in particular interesting for the exceptional ALE singularities, where a brane realization in Type I$'$ is lacking. Our approach to determine these models is based on 6d conformal matter: we determine these theories as generalized 6d quivers.  All these LSTs have a higher-one form symmetry which forms a 2-group with the zero-form Poincar\'e symmetry, the R-symmetry and the other global symmetries: the matching of the R-symmetry two-group structure constant is a stringent constraint for T-dualities, which we use in combination with the matching of 5d Coulomb branches and flavor symmetries upon circle reduction, as a consistency check for the realization of the 6d LSTs we propose.
\end{abstract}

\vfill{}
--------------------------

\today

\thispagestyle{empty}

\newpage

\tableofcontents

\section{Introduction}

This paper is the first of a series devoted to revisiting the properties of the Heterotic strings on ALE spaces. The main motivation for our study is that recent progress in understanding the structures of six-dimensional theories \cite{Heckman:2013pva,DelZotto:2014hpa,Ohmori:2014kda,DelZotto:2014fia,DelZotto:2015rca,Heckman:2015bfa,Ohmori:2015pua,Ohmori:2015pia,Bhardwaj:2015oru,DelZotto:2017pti,Mekareeya:2017jgc,Ohmori:2018ona,Dierigl:2020myk} and their continuous 2-group symmetries \cite{DelZotto:2015isa,Cordova:2018cvg,Cordova:2020tij,Bhardwaj:2020phs,DelZotto:2020sop,Apruzzi:2021mlh} can be exploited to give new insights on some of the open questions in the subject.

\medskip

The focus of this work are the little string theories (LSTs) of Heterotic ALE instantons and the corresponding T-dualities \cite{Aspinwall:1996vc,Aspinwall:1997ye,Blum:1997mm,Intriligator:1997dh,Hanany:1997gh,Brunner:1997gf}. The LSTs governing the worldvolumes of the heterotic $Spin(32)/\mathbb Z_2$ ALE instantons are obtained from orbifolds of the Lagrangian theory governing a stack of N NS5 branes of the $Spin(32)/\mathbb Z_2$ Heterotic string and are well-known \cite{Sagnotti:1987tw,Bianchi:1990yu,Blum:1997mm,Intriligator:1997dh}. On the contrary, the LSTs governing the worldvolumes of the Heterotic $E_8 \times E_8$ ALE instantons are non-Lagrangian and slightly more mysterious: our first result in this context is to completely determine the latter exploiting 6d conformal matter, thus extending previous results in the literature \cite{Aspinwall:1997ye,Font:2017cya,Font:2017cmx}. To confirm our predictions we exploit T-duality with the known $Spin(32)/\mathbb Z_2$ ALE instantons LSTs. The main criteria we use for identifying T-dual pairs of theories are the matching of flavor symmetry ranks, 5d Coulomb branch dimensions, and 2-group structure constants \cite{Cordova:2020tij,DelZotto:2020sop}. The latter constraint is often the most stringent, and allows to chart the corresponding T-dual models purely from a field theoretical perspective. We conjecture that the matching of the above data is sufficient to predict a T-duality between a pair of LSTs. As a consequence we end up predicting several new families of equivalences among these models.



\medskip

As a further consistency checks of the results above we exploit brane constructions in Type I and Type I$'$ for some of the models of interest. These brane engineerings however are not effective for exceptional ALE singularities. For those cases, we can confirm our results by exploiting the geometrization of T-duality in F-theory \cite{Aspinwall:1997ye,DelZotto:2015rca}. The detailed analysis of the relevant geometries will appear in a follow up work in this series \cite{DZLO3}.

\medskip

The structure of this paper is as follows. In section \ref{sec:2groupreview} we establish the notations and conventions used throughout this paper and we review the relevant aspects of the 2-group global symmetry of 6d LSTs. In section \ref{sec:tenzorb} we determine all the 6d LSTs for the Heterotic ALE  $E_8\times E_8$ instantons in presence of arbitrary choices of flat connections at infinity. In section \ref{sec:e8Ak} we describe in details the case of $A_k$-type singularities, exploiting the duality with Type I$'$ superstrings to fully chart the T-duality landscape for some small values of $k$. In section \ref{sec:duality3} we briefly discuss some aspects of the case of $D$-type singularities. In section \ref{exceptional} we discuss the case of exceptional singularities, with particular focus on the choices of flat connections at infinity which give rise to exceptional flavor symmetries. We focus on these examples for the sake of brevity, but our methods are valid in full generality. We conclude in section \ref{sec:atheorem} presenting a conjecture about the role of the $R$-symmetry 2-group structure constant to constrain RG flows. By direct inspection of the cases considered in this paper, we see that indeed the latter is decreasing along RG flows.

\medskip

\noindent \textbf{Note added.} \textit{While this work was in preparation we have been informed about \cite{Bhardwaj:2022ekc} which obtained some of the results we present here in the context of T-duality with different methods. We thank the author of that manuscript for coordinating the publication.}

\section{A quick review of 2-groups for LSTs}\label{sec:2groupreview}

In this section we fix our notations and conventions for generalized quiver diagrams \cite{DelZotto:2014hpa}  for 6d LSTs \cite{Bhardwaj:2015oru} and we illustrate the formulas for the 2-group structure constants obtained in \cite{Cordova:2020tij,DelZotto:2020sop}.\footnote{\ We stress this is by no means meant to be a comprehensive review about the physics of these models: we refer the readers interested in a review to section 2 of \cite{DelZotto:2018tcj} or to the manuscript \cite{Heckman:2018jxk}.} Experienced readers can safely skip to the next section.

\medskip

Consider a 6d LSTs of rank $n_T$, with a global 6d zero-form symmetry with Lie algebra
\be
\mathfrak f^{(0)} = \prod_{a=1}^{n_f} \mathfrak f_a\,,
\ee
where $\mathfrak f_a$ are irreducible factors. Its generalized quiver is encoded by two sets of data:
\begin{itemize}
\item An $(n_T+1 + n_f) \times (n_T+1 + n_f)$ symmetric matrix 
\be
\left(\begin{matrix} \eta^{IJ} & \eta^{IA} \\ \eta^{AI} & 0\end{matrix}\right) \qquad \begin{aligned}& I,J = 1,...,n_T+1 \\ &A = 1,...,n_f \end{aligned}
\ee
\item A $(n_T+1 + n_f)$-touple of Lie algebras
\be
\mathfrak g = (\mathfrak g_1,..., \mathfrak g_{n_T+1}, \mathfrak f_1,..., \mathfrak f_{n_f})\,,
\ee
\end{itemize}
 One associates a node of the generalized quiver to each $1\leq I \leq n_T+1$, decorated with the value of $\eta^{II}$, the Lie algebra $\mathfrak g_I$, and, whenever $\eta^{IA} \neq 0$, a factor $\mathfrak f_A$ of the symmetry Lie algebra in square brackets, schematically
\be
\cdots \, \underset{\atop [\mathfrak f_A]_{(\eta^{\text{\tiny{\textsc{ia}}}})}}{\overset{\mathfrak g_I \atop \,}{\eta^{II}}} \, \cdots
\ee
The algebras $\mathfrak g_I$, $I=1,...,n_T+1$ above correspond to dynamical gauge fields. The diagonal entries of the $\eta^{IJ}$ block are positive integers between 0 and 12 encoding the self Dirac pairings for the $n_T+1$ elementary BPS strings of the theory, which source the corresponding selfdual 2-form gauge fields $b^{(2)}_I$. The off-diagonal entries of the $\eta^{IJ}$ block are non-positive integers encoding the adjacency matrix for the generalized quiver: if $\eta^{IJ} \neq 0$ the two nodes $\eta^{II}$ and $\eta^{JJ}$ are adjacent and the corresponding BPS strings can form bound-states. For all cases we consider in this paper, the off-diagonal entries of $\eta^{IJ}$ are either $0$ or $-1$, but more general cases are indeed possible \cite{Bhardwaj:2015oru,Bhardwaj:2018jgp}. The decoration by $\mathfrak g_I$ is typically suppressed for those nodes with $\eta^{II}=1$ and $\mathfrak g_I = \mathfrak{sp}_0$ (for which $h^\vee \equiv 1$), and $\eta^{II}=2$ with an $\mathfrak g_I=\mathfrak{su}(1)$ (for which also $h^\vee \equiv 1$).  Matter in representations to cancel the corresponding quartic gauge anomalies is typically added, but often is suppressed in the notation, together with the coefficient $\eta^{IA}$, which is determined by it. For more details about the physics interpretation of this notation, we refer our readers to the papers \cite{DelZotto:2014hpa,Heckman:2015bfa}, as well as to \cite{Bhardwaj:2015oru} for its application to LSTs.

\medskip

For 6d SCFTs the matrix $\eta^{IJ}$ must be postive definite, while for 6d LSTs it has to be non-negative. The little string of the theory is given by a boundstate of elementary BPS strings with charge encoded in the unique null eigenvector of $\eta^{IJ}$ (which is a generalized \emph{affine} Cartan matrix) \cite{Bhardwaj:2015oru}:
\be\label{eq:lstcharge}
\eta^{IJ} \ell_J = 0 \qquad \gcd(\ell_1,...,\ell_{n_T+1}) = 1 \qquad \ell_I > 0\,.
\ee
Corresponding to this charge, a feature of 6d LSTs is that the following linear combination of 2-form tensor fields is background
\be
B^{(2)}_\textsc{lst} = \sum_{I=1}^{n_T+1} \ell_I b^{(2)}_I
\ee
and corresponds to a 1-form symmetry $U(1)^{(1)}_\textsc{lst}$. The associated background curvature 3-form $H^{(3)}_{\textsc{lst}}$ satisfies a modified Bianchi identity involving the background instanton densities, which, if non-trivial, control the 2-group structure constants $\hat\kappa_F$, $\hat\kappa_{\mathscr P}$ and $\hat\kappa_R$  \cite{Cordova:2020tij}:
\be\label{eq:structure}
\begin{aligned}
\frac1{2\pi} d H^{(3)}_{\textsc{lst}} &= \frac1{2\pi} \sum_{I=1}^{r+1}\ell_I dH_I^{(3)}\\&= \hat \kappa_R c_2(R) - \frac{\hat \kappa_\mathscr{P}}{4} p_1(TM_6)-\frac14 \hat \kappa_F \mathrm{Tr} F_A^2\,.
\end{aligned}
\ee
In presence of nontrivial backgrounds for the zero-form global symmetries of the 6d theory, the two-form fields $b^{(2)}_I$ have Green-Schwartz couplings of the form \cite{Ohmori:2014kda}
\begin{equation}\label{eq:GSCUP}
\int_6 b^{(2)}_I \wedge \left(h^\vee_{\mathfrak{g}_I} c_2(R) + \frac14 (\eta^{II}-2) p_1(TM_6)+\frac14\eta^{IA}\mathrm{Tr}(F_A^2)\right)\,,
\end{equation}
where $\eta^{II}$ is not summed over, $c_2(R)$ and $p_1(TM_6)$ correspond to backgrounds for the $Sp(1)_R$ symmetry of the theory and gravity respectively, and $F_A$ is a background field strength for the $A$-th factor of the global flavor symmetry group. In presence of the GS couplings in equation \eqref{eq:GSCUP}, all the tensor fields have modified Bianchi identities of the
form 
\be
 \frac1{2\pi} d H^{(3)}_{I} = h^\vee_{\mathfrak{g}_I} c_2(R) + \frac14 (\eta^{II}-2) p_1(TM_6)+\frac14\eta^{IA}\mathrm{Tr}(F_A^2),
\ee
Plugging this equation into \eqref{eq:structure}, gives an equation for the 2-group structure constants of the model
   \be\label{eq:2grstr}
   \framebox{$\phantom{\Bigg|}\hat \kappa_F =  - \sum_{I=1}^{r+1} \ell_I \eta^{IA} \qquad \hat \kappa_R =   \sum_{I=1}^{r+1} \ell_I h^\vee_{\mathfrak{g}_I} \qquad \hat \kappa_\mathscr{P} = - \sum_{I=1}^{r+1} \ell_I (\eta^{II}-2)\,.$}
   \ee
In reference \cite{DelZotto:2020sop} it was checked that the above 2-group structure constants always match for T-dual LSTs. The purpose of this work is to exploit such matching to chart the possible T-dualities among Heterotic ALE instantonic LSTs.

\section{The LSTs of Heterotic $E_8 \times E_8$ ALE instantons}\label{sec:tenzorb}

\begin{figure}
\begin{center}
\includegraphics[scale=0.7]{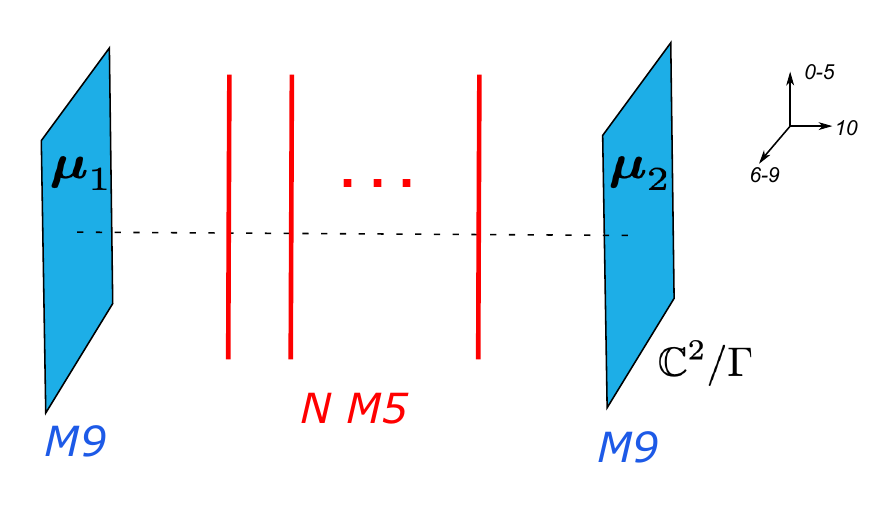}
\caption{Exceptional LSTs in the Hořava-Witten duality frame.}\label{fig:HWframe}
\end{center}
\end{figure}

The worldvolume theories governing a stack of $N$ instantonic NS5 branes of type $Spin(32)/\mathbb Z_2$ are well-known \cite{Witten:1995gx}: they have a Lagrangian description in terms of an $\mathfrak{sp}(N)$ gauge theory coupled to $N_f = 16$ hypermultiplets in the fundamental representation, as well as one antisymmetric. The corresponding generalized quiver is
\be
[\mathfrak{so}(32)] \overset{\mathfrak{sp}_N}{0} [\mathfrak{su}(2)]\,.
\ee
Thanks to this a Lagrangian, the worldvolume theories of ALE heterotic $Spin(32)/\mathbb Z_2$ can be easily determined by orbifolding \cite{Blum:1997fw,Blum:1997mm,Intriligator:1997dh}. Since $\pi_1(S^3/\Gamma_{\mathfrak g}) \simeq \Gamma_{\mathfrak g}$, the resulting theories also depend on the choice of a flat connection at infinity, encoded in a choice of a mapping 
\be
\boldsymbol{\lambda}:\Gamma_{\mathfrak{g}} \to Spin(32)/\mathbb Z_2.
\ee
We review the structure of the relevant models when needed in the analysis below. We denote them
\be
\widetilde{\mathcal K}_N(\boldsymbol{\lambda}; \mathfrak g)
\ee
in what follows. These models have 6d flavor symmetry $F(\boldsymbol{\lambda})$ which is determined by the commutant of $\boldsymbol{\lambda}(\Gamma_{\mathfrak g})$ in $Spin(32)/\mathbb Z_2$.

\medskip

An important subtlety here is that one should distinguish whether these instantons give obstructions to a ``vector structure'' for $Spin(32)/\mathbb Z_2$ via a positive second Stieffel-Whitney class $\widetilde{\textsf{w}}_2 \neq 0$ or not -- as remarked in \cite{Berkooz:1996iz}. In this paper, as well as in parts 2 and 3, we will mostly focus on the case in which $\widetilde{\textsf{w}}_2 = 0$, since, as we will discuss below, we expect the cases with $\widetilde{\textsf{w}}_2 \neq 0$ to be dual to configurations in the frozen phase of F-theory, which is still relatively unexplored \cite{Bhardwaj:2018jgp,Bhardwaj:2019hhd}.\footnote{\ We plan to return to this issue in part IV of this project, which is currently under preparation.}

\medskip

On the contrary, the theories associated to instantons on $E_8 \times E_8$ are close cousins of the 6d (2,0) SCFTs, and in particular one does not expect these models to have a simple Lagrangian formulation. A beautiful characterization for the LSTs of fractional $E_8 \times E_8$ heterotic instantons is achieved via the Hořava-Witten duality between Heterotic superstrings and M-theory on a finite interval \cite{Horava:1995qa}. In the M-theory dual frame, the exceptional LSTs arise from a stack of $N$ M5 branes extended along the directions $x_{0,1,2,3,4,5}$ that are parallel to the two M9 branes at the opposite ends of the world \cite{Ganor:1996mu}. Along the directions $x_{6,7,8,9}$ transverse to the M5 branes, an ALE singuarity $\mathbb C^2/\Gamma_{\mathfrak g}$ is located -- see Figure \ref{fig:HWframe}. Due to the presence of the singularity, the instantonic configuration fractions and the resulting theory depends on a further choice of a flat connection at infinity for the two $E_8$ bundles. The latter are encoded in two group morphisms 
\be
\boldsymbol{\mu}_{a} \colon \pi_1(S^3/\Gamma_{\mathfrak g}) \simeq \Gamma_{\mathfrak g}\to E_8\,,
\ee
where $a=1,2$ is a label for the two M9 branes. We represent this graphically in Figure \ref{fig:HWframe} as a decoration of the M9 brane. The zero form global symmetry of the resulting LSTs is determined by the commutant of $\boldsymbol{\mu}_a(\Gamma_{\mathfrak g})$ in $E_8$, namely we expect to have 
\be
F^{(0)}_a \equiv \{ g \in E_8 \, | \, gh = hg, \forall h \in \boldsymbol{\mu}_a(\Gamma_{\mathfrak g})\} \qquad a=1,2\,. \\
\ee
The cases with global symmetry $F_{1,2}^{(0)} \simeq E_8$, corresponding to the choice of trivial flat connections for the end of the world gauge fields, are dual to certain geometries in F-theory, discovered by Aspinwall and Morrison \cite{Aspinwall:1997ye}. In this paper, we are concerned with all other possible choices. We denote the corresponding theories
\be
\mathcal K_N(\boldsymbol{\mu}_1,\boldsymbol{\mu}_2;\mathfrak g)
\ee

\medskip

The precise form of the tensor branches for all these classes of theories can be easily obtained from the conformal matter approach. The distance M9-M5 corresponds to the vev of a (1,0) tensormultiplet associated to a BPS string with unit self Dirac pairing, while the distance M5-M5 corresponds to the vev of a (1,0) tensormultiplet associated to a BPS string with self Dirac pairing two, leading to a structure
\be
1\,\underbrace{2\,2\,2\, \cdots \, 2 \, 2}_{N-1}\,1
\ee
In presence of a $\mathbb C^2/\Gamma_{\mathfrak g}$ singularity the various branes involved can fraction. The resulting fractions have been determined exploiting F-theory techniques \cite{DelZotto:2014hpa} -- for an application in the context of this paper, we refer our readers to \cite{DZLO3}. 
The resulting theories are described as generalized quiver theories of the form
\be\label{eq:avheterotto}
\mathcal K_N(\boldsymbol{\mu}_1,\boldsymbol{\mu}_2;\mathfrak g) = \xymatrix{\mathcal T(\boldsymbol{\mu_1},\mathfrak g) \ar@{-}[r]^{\mathfrak {g}} &\mathcal T_{N-2}(\mathfrak g,\mathfrak g)  \ar@{-}[r]^{\mathfrak {g}}&\mathcal T(\boldsymbol{\mu_2},\mathfrak g) }
\ee
where:
\begin{itemize}
\item $\mathcal T(\boldsymbol{\mu_a},\mathfrak g)$ is the minimal 6d orbi-instanton theory associated to the M9-M5 system in presence of a $\mathbb C^2/\Gamma_{\mathfrak g}$ transverse to the M5, with a choice of flat connection at infinity $\boldsymbol{\mu_a}:\Gamma_{\mathfrak g} \to E_8$;
\item $\mathcal T_{N-2}(\mathfrak g,\mathfrak g)$ is the 6d conformal matter theory associated to $N-2$ M5 branes probing a $\mathbb C^2/\Gamma_{\mathfrak g}$ singularity;
\item $\xymatrix{\ar@{-}[r]^{\mathfrak g}&}$ denotes the operation of (diagonal) fusion of the common factors $\mathfrak g$ of the global symmetry of the corresponding 6d SCFTs, schematically at the level of the corresponding generalized quivers:\footnote{\ This is the 6d version of the gauging operation in 4d, which our readers are probably more familiar with. For further references about this, see \cite{DelZotto:2018tcj} and \cite{Heckman:2018pqx}. Our readers that are not familiar with this operation can find plenty of examples in the discussion below.}
\be
\xymatrix{\cdots \,\, \overset{\mathfrak g'}{n'} \,\, \textcolor{red}{[\mathfrak g]} \ar@[red]@{-}[r]^{\textcolor{red}{\mathfrak g}}\,\,& \textcolor{red}{[\mathfrak g]}   \,\,  \overset{\mathfrak g''}{n''} \cdots }\,\, \longrightarrow \,\,\cdots \overset{\mathfrak g'}{n'} \,\,\textcolor{red}{\overset{\mathfrak{g}}{n}} \,\, \overset{\mathfrak g''}{n''} \cdots
\ee
\end{itemize}
The theories $\mathcal T(\boldsymbol{\mu_a},\mathfrak g)$ can be determined from results found in references  \cite{DelZotto:2014hpa,Heckman:2015bfa,Mekareeya:2017jgc,Frey:2018vpw}, by identifying the minimal orbi-instanton model, corresponding to a single M5-M9 system in presence of a transverse ALE singularity. When $N \geq 3$, the structure we describe in \eqref{eq:avheterotto} completely determines the tensor branches of all possible fractional $E_8 \times E_8$ heterotic instantons for all possible singularities, thus nicely complementing the results available for this class of models in the literature \cite{Aspinwall:1996vc,Aspinwall:1997ye,Intriligator:1997dh}.

\medskip

\noindent \textbf{Remarks:}
\begin{enumerate}
\item We stress here that the $N\leq 2$ cases deviate slightly from the structure above. The theories corresponding to such small $N$ cases are analysed in details in the second paper of this series \cite{DZLO2}, where an application of these methods to determine the geometric engineering limits of the Heterotic Strings on ALE singularities is also presented.
\item When indicating the flavor symmetries below we will not be careful about the global form of the group, which are inessential for the main purpose of this note.
\item From the structure of the theories above, we see that all these theories will have $\widehat{\kappa}_{\mathscr{P}} = 2$. The most interesting 2-group structure constant is
\be
\framebox{$\phantom{\Big|}\widehat{\kappa}_R = \widehat{\kappa}_{\mathfrak{g},N} + \widehat{\kappa}_{\mathfrak{g},\boldsymbol{\mu}_1} + \widehat{\kappa}_{\mathfrak{g},\boldsymbol{\mu}_2}$}
\ee
where
\begin{itemize}
\item $\widehat{\kappa}_{\mathfrak{g},N}$ is the contribution to $\widehat{\kappa}_R$ coming from the conformal matter of type $\mathcal{T}_{N-2}(\mathfrak{g},\mathfrak{g})$ with the addition of the contribution from the two gauge groups involved in the fission procedure;
\item $\widehat{\kappa}_{\mathfrak{g},\boldsymbol{\mu}_a}$ is the contribution to $\widehat{\kappa}_R$ arising from the models $\mathcal{T}(\boldsymbol{\mu}_a,\mathfrak{g})$
\end{itemize}
These quantities are additive and have the structure above because the LS charge 
\be
\eta \cdot \vec{\ell}_{LS} = 0
\ee
factors along the genrealized quiver with structure
\be
\vec{\ell}_{LS} = (\vec{\ell}_{\mathfrak{g},\boldsymbol{\mu}_1},\ell_{\mathfrak{g},1},\vec{\ell}_{(N,\mathfrak{g})},\ell_{\mathfrak{g},2},\vec{\ell}_{\mathfrak{g},\boldsymbol{\mu}_2}) 
\ee
compatible with the decomposition in equation \eqref{eq:avheterotto} above. Then we have that
\be
\widehat{\kappa}_{\mathfrak g,\boldsymbol{\mu}_a} = \vec{\ell}_{\mathfrak{g},\boldsymbol{\mu}_a} \cdot \vec{h}^\vee(\mathcal T(\boldsymbol{\mu}_a,\mathfrak g))
\ee
Moreover,
\be
\widehat{\kappa}_{\mathfrak{g},N} = (\ell_{\mathfrak g,1} + \ell_{\mathfrak g,2}) h^\vee_{\mathfrak g} + \vec{\ell}_{(N,\mathfrak{g})} \cdot \vec{h}^\vee(\mathcal T_{N-2}(\mathfrak g,\mathfrak g))
\ee
where $\ell_{\mathfrak g,a}$ is the coefficient of the LS charge corresponding to the $a$-th fusion node, and we have introduced the notation $\vec{h}^\vee(\mathcal T)$: For a 6d theory $\mathcal T$ with generalized quiver
\be
\overset{\mathfrak g_1}{n_1}\cdots \overset{\mathfrak g_I}{n_I}\cdots \overset{\mathfrak g_r}{n_r}
\ee
we define 
\be
\vec{h}^\vee(\mathcal T) \equiv (h^\vee_{\mathfrak{g}_1},\dots, h^\vee_{\mathfrak{g}_I}, \dots, h^\vee_{\mathfrak{g}_r})\,.
\ee

\item Sometimes it can happen that a pair of models $\mathcal T(\boldsymbol{\mu},\mathfrak g)$ and $\mathcal T(\boldsymbol{\mu}',\mathfrak g)$ have
\be
\widehat{\kappa}_{\mathfrak{g},\boldsymbol{\mu}} = \widehat{\kappa}_{\mathfrak{g},\boldsymbol{\mu}'}\,.
\ee
For all those examples the theories
\be
\mathcal K_N(\boldsymbol{\mu},\boldsymbol{\mu}''; \mathfrak g) \qquad\text{and}\qquad \mathcal K_N(\boldsymbol{\mu}',\boldsymbol{\mu}''; \mathfrak g)
\ee
will have the same $\widehat{\kappa}_R$.
\end{enumerate}
Given the above data the question we are addressing in this paper is to chart the T-dualities
\be
\widetilde{\mathcal K}_{\widetilde{N}}(\boldsymbol{\lambda};\mathfrak g) \sim \mathcal K_N(\boldsymbol{\mu}_1,\boldsymbol{\mu}_2;\mathfrak g)\,.
\ee
For a pair of models to be T-dual, the following conditions must be met
\begin{itemize}
\item The flavor symmetry ranks must match: $$\text{rk }F(\boldsymbol{\lambda}) = \text{rk }F(\boldsymbol{\mu}_1)+ \text{rk }F(\boldsymbol{\mu}_2)\,= \text{rk }F_{5d} - 1$$
where we have subtracted the contribution of the KK charge from $\text{rk }F_{5d}$;
\item The dimensions of the Coulomb branches of the 5d theories obtained by the circle reduction of the two models must match;
\item The 2-group structure constants must match across T-duality. For all these models
\be
\widehat{\kappa}_{\mathscr{P}} = 2 \, .
\ee
A much stronger constraint is provided by the requirement that 
\be
\widehat{\kappa}_{R}(\widetilde{\mathcal K}_{N}(\boldsymbol{\lambda};\mathfrak g) ) = \widehat{\kappa}_{R}(\mathcal K_N(\boldsymbol{\mu}_1,\boldsymbol{\mu}_2;\mathfrak g)) \, .
\ee
Moreover, of course, one has also to check that the 2-group structure constants corresponding to the 6d flavor symmetries of these models do indeed match. In order to do that one has to remember that often along the T-duality circles one can introduce Wilson lines breaking the two flavor symmetries to maximal subalgebras
\be
\xymatrix{F(\boldsymbol{\lambda}) \ar[dr]^{\textbf{W}_{\boldsymbol{\lambda}}}\\
&F_{5d}\\
 F(\boldsymbol{\mu}_1) \times F(\boldsymbol{\mu}_2)\ar[ur]^{\textbf{W}_{\boldsymbol{\mu}_1} \oplus \textbf{W}_{\boldsymbol{\mu}_2}\quad}}
\ee
Then the 2-group structure constants can be compared in 5d as they correspond to the same $F_{5d}$. Along the process one might have to rescale them according to the index of embedding of $F_{5d}$ in the 6d flavor symmetry groups. In all the cases we consider in this paper such an index equals one and the flavor structure constants are easily matched, we therefore omit them from our tables.
\end{itemize}

The requirements above are necessary for a pair of LSTs to be T-dual. For the examples we consider in this paper, we conjecture these requirements are also sufficient. Evidence for this conjecture is obtained exploiting Type I$'$ geometric engineering for these systems and the behavior of membranes upon string dualities, which give the Type I$'$ version of the Heterotic T-dualities. For the case of exceptional singularities, no brane engineering is available and one has to turn to F-theory for checking these conjectures. This is the subject of Part III of this series of works \cite{DZLO3}.

\section{Heterotic instantons on $\mathbb C^2/\mathbb Z_k$ singularities}\label{sec:e8Ak}

In this section we review the results for $\mathfrak g =\mathfrak a_{k-1}$ singularities, $\mathbb C^2 / \mathbb Z_k$. In this case the possible $\boldsymbol{\mu}: \mathbb Z_k \to E_8$ are classified by a simple rule \cite{KAC} (see also Section 7 of \cite{Heckman:2015bfa}). Each different $\boldsymbol{\mu}$ corresponds to a decomposition of $k$ into a sum (with repetitions) of the form
\be\label{eq:E8Ak}
k = \sum_{j} d_{i_j}
\ee
where $d_i$ are the Kac labels for $E_8$, positive integers corresponding to the nodes of the $\widehat{E}_8$ diagram as follows
\be\label{eq:E8KAC}\xymatrix@=0.8pc{&&&&&\overset{3'}{\bullet}\ar@{-}[d]\\
\overset{1}{\circ} \ar@{-}[r]&\overset{2}{\bullet} \ar@{-}[r]&\overset{3}{\bullet} \ar@{-}[r]&\overset{4}{\bullet} \ar@{-}[r]&\overset{5}{\bullet} \ar@{-}[r]&\overset{6}{\bullet} \ar@{-}[r]&\overset{4'}{\bullet}\ar@{-}[r]&\overset{2'}{\bullet}}
\ee

\medskip

The corresponding maximal subgroup $F$ of $E_8$, which commutes with such an embedding has the Dynkin diagram obtained by deleting from the the diagram in equation \eqref{eq:E8KAC} the nodes corresponding to the $d_i$'s which enter in the decomposition in \eqref{eq:E8Ak}. A nice algorithm which determines all possible $\mathcal T(\boldsymbol{\mu},\mathfrak{su}_{k})$ theories can be found in reference \cite{Mekareeya:2017jgc}. 

\subsection{Type I$'$ formulation}\label{sec:duality1}

When the singularities involved are of type $\mathfrak g = \mathfrak a_{k-1}$ or $\mathfrak d_k$, one can consider dual Type I$'$ configurations, which we can take advantage of in order to track T-dualities. In what follows we consider the $\mathfrak a_{k-1}$ cases and we review some aspects of the relevant dualities following the discussion in \cite{Horava:1995qa,Polchinski:1995df} --- see also \cite{Intriligator:1997dh,Brunner:1997gf,Hanany:1997gh}.

\medskip

\begin{figure}
\begin{center}
\includegraphics[scale=0.5]{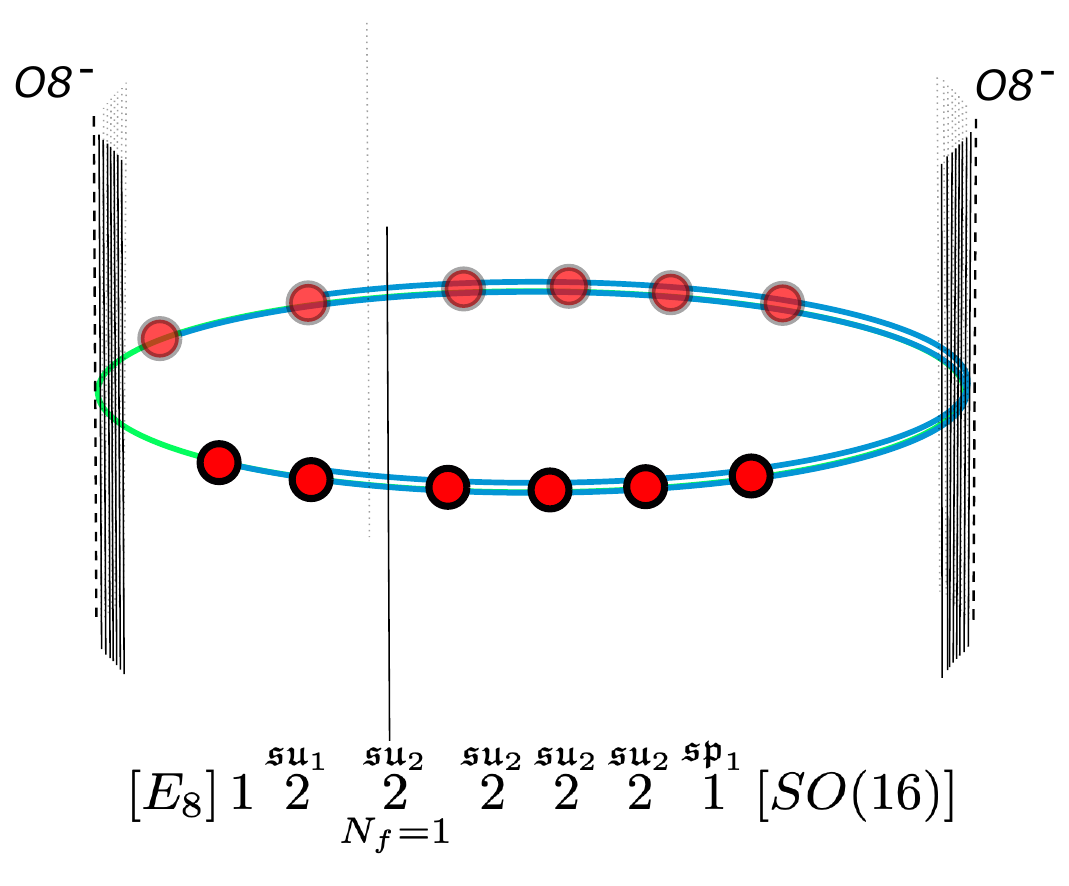}
\caption{Example of Type I$'$ dual configuration to a Hořava-Witten setup: here we have instantons probing a $\mathbb C^2/\mathbb Z_2$ singularity with $\boldsymbol{\mu}_1 = 1+1$, $\boldsymbol{\mu}_2 = 2'$ }\label{fig:E8typeI}
\end{center}
\end{figure}

Let us proceed by dualizing the Hořava-Witten setup to a Type I$'$ system \cite{Polchinski:1995df}. In order to do that it is convenient to realize the $\mathbb C^2/\mathbb Z_k$ singularity as a charge $k$ Taub-NUT space $\mathbf{TN}_k$ and then use the Taub-NUT circle as an M-theory circle, which morphs the Taub-NUT metric into a stack of $k$ D6 branes \cite{Sen:1997js}. It is convenient to summarize schematically the configuration as follows (see also Figure \ref{fig:HWframe})
\be\label{eq:HW}
\begin{tabular}{c|ccccccccccc}
&0 & 1 & 2 & 3 & 4 & 5 & 6 & 7 & 8 & 9 & 10 \\
\hline
M9 & $\bullet$ & $\bullet$ & $\bullet$ & $\bullet$ & $\bullet$& $\bullet$& $\bullet$& $\bullet$& $\bullet$& $\bullet$ & \\ 
$N$ M5 & $\bullet$ & $\bullet$ & $\bullet$ & $\bullet$ & $\bullet$ & $\bullet$ \\
$\mathbf{TN}_k$ & $\bullet$ & $\bullet$ & $\bullet$ & $\bullet$ & $\bullet$ & $\bullet$ & $\circ$ & $\circ$ & $\circ$ & $\circ$ &  $\bullet$ \\
\end{tabular}
\ee
Dualizing, one obtains the following IIA brane system
\be\label{eq:IIAbrain}
\begin{tabular}{c|ccccccccccc}
&0 & 1 & 2 & 3 & 4 & 5 & 6 & 7 & 8  & 10\\
\hline
O8-D8 & $\bullet$ & $\bullet$ & $\bullet$ & $\bullet$ & $\bullet$& $\bullet$& $\bullet$& $\bullet$& $\bullet$&  & \\ 
$M$ NS5 & $\bullet$ & $\bullet$ & $\bullet$ & $\bullet$ & $\bullet$ & $\bullet$ \\
$k$ D6 & $\bullet$ & $\bullet$ & $\bullet$ & $\bullet$ & $\bullet$ & $\bullet$ & & & &  $\bullet$ \\
\end{tabular}
\ee
Recall that $x^{10}$ is an interval of the form $S^1/\mathbb Z_2$. The two M9 branes are mapped to two $O8^-$ planes located at the antipodal points each associated with $8$ D8 branes (and their images) so that the overall Roman's mass of the configuration is zero. The rules to manipulate these diagrams are well known (see e.g. \cite{Hanany:1997gh,Brunner:1997gf}). The $\mathbb C^2/\mathbb Z_k$ singularity becomes a stack of $k$ D6 branes which are wrapping around the $S^1/\mathbb Z_2$. Naively one might say that the $N$ M5 branes are mapped to $N$ NS5 branes, but in facts this is not the case, which is due to the fact that the M9 fractions along the singularity \cite{DelZotto:2014hpa}. We instead obtain a total of
\be
M=N + N_{\boldsymbol{\mu}_1} + N_{\boldsymbol{\mu}_2}
\ee
NS5 branes in this case: part of the NS5 branes in the dual Type I$'$ configuration are dual to fractions of M9 branes, which crucially depend also on the choice of $\boldsymbol{\mu}_a$, that, moreover, encodes the position of the $D8$ branes relative to the NS5s. For a simple example our readers can look at Figure \ref{fig:E8typeI}, where we see that for $\boldsymbol{\mu}_2 = 2'$ (on the right) the M9 does not fraction, while for $\boldsymbol{\mu}_1 = 1+1$ (on the left), the corresponding M9 indeed fractions.

\medskip

Now, since we are interested in T-duality we can add an extra circle within the M5 branes worldvolume. Let's choose the $x^5$ coordinate in the Hořava-Witten setup to be such an $S^1$. Then we could use that circle as an M-theory circle which gives the IIA brane system
\be
\begin{tabular}{c|ccccccccccc}
&0 & 1 & 2 & 3 & 4 & 6 & 7 & 8 & 9 & 10 \\
\hline
O8-D8 & $\bullet$ & $\bullet$ & $\bullet$ & $\bullet$& $\bullet$& $\bullet$& $\bullet$& $\bullet$& $\bullet$ & \\ 
$N$ D4 & $\bullet$ & $\bullet$ & $\bullet$ & $\bullet$ & $\bullet$ \\
$\mathbf{TN}_k$ & $\bullet$ & $\bullet$ & $\bullet$ & $\bullet$ & $\bullet$ & $\circ$ & $\circ$ & $\circ$ & $\circ$ &  $\bullet$ \\
\end{tabular}
\ee
At this point we can dualize to IIB along the Taub-NUT circle to obtain
\be\label{eq:IIB}
\begin{tabular}{c|ccccccccccc}
&0 & 1 & 2 & 3 & 4 & 6 & 7 & 8 & 9$'$ & 10 \\
\hline
O7-D7 & $\bullet$ & $\bullet$ & $\bullet$ & $\bullet$& $\bullet$& $\bullet$& $\bullet$& $\bullet$&  & \\ 
$N$ D5 & $\bullet$ & $\bullet$ & $\bullet$ & $\bullet$ & $\bullet$ &&&& $\bullet$ \\
$k$ NS5 & $\bullet$ & $\bullet$ & $\bullet$ & $\bullet$ & $\bullet$ & &&  &  &  $\bullet$ \\
\end{tabular}
\ee
which is a configuration that we can uplift back to Type I$'$ by T-dualizing along the 10-th direction, thus giving:
\be
\begin{tabular}{c|ccccccccccc}
&0 & 1 & 2 & 3 & 4 & 6 & 7 & 8 & 9$'$ & 10$'$ \\
\hline
O8-D8 & $\bullet$ & $\bullet$ & $\bullet$ & $\bullet$& $\bullet$& $\bullet$& $\bullet$& $\bullet$&  & $\bullet$\\ 
$N$ D6 & $\bullet$ & $\bullet$ & $\bullet$ & $\bullet$ & $\bullet$ &&&& $\bullet$&$\bullet$ \\
$k$ NS5 & $\bullet$ & $\bullet$ & $\bullet$ & $\bullet$ & $\bullet$ & &&  &  &  $\bullet$ \\
\end{tabular}
\ee
The latter is a Type I$'$ configuration that has an interpretation as the LST of $N$ Heterotic $Spin(32)/\mathbb Z_2$ instantons probing a $\mathbb C_2/\mathbb Z_k$ after \cite{Intriligator:1997dh,Brunner:1997gf,Hanany:1997gh}. In the process of T-dualising along the 10-th direction, the orientifold planes are merged together and recombine, which signals non-perturbative effects kick-in from the Heterotic perspective. Here the group morphism
\be
\boldsymbol{\lambda}:\mathbb Z_k \to Spin(32)/\mathbb Z_2
\ee
is encoded by the relative position of the 16 D8s with respect to the NS5s.

\medskip

Exploiting the matching of the corresponding 2-group structures and 5d Coulomb branch dimensions we can proceed charting in details the corresponding T-duals
\be
\mathcal K_N(\boldsymbol{\mu}_1,\boldsymbol{\mu}_2; \mathfrak g = \mathfrak a_{k-1}) \, \leftrightarrow \, \widetilde{\mathcal K}_N(\boldsymbol{\lambda}; \mathfrak g = \mathfrak a_{k-1}) 
\ee
In this context it is interesting to see how the possible choices of $\boldsymbol{\mu}_1$ and $\boldsymbol{\mu}_2$ are mapped to choices of $\boldsymbol{\lambda}$.

\medskip

In all these examples the generalized quiver diagrams have $\eta^{IJ}$ of the form
\be
\eta^{IJ} = \left(\begin{matrix}1 & -1 &  0 & 0 & \cdots & 0 \\ -1 & 2 & -1 & 0 & \cdots  & 0\\ \vdots& &&&& \vdots\\ & & \cdots & -1 &2 & -1 \\ & &  \cdots & 0 &-1 & 1 \\ \end{matrix}\right)
\ee
where the fact that there are two BPS strings with charge 1 ultimately follows from the presence of the two O8$^-$ planes. The corresponding LS charge is
\be
\vec{\ell}_{LS}=(1,1,\dots,1)
\ee
which simplifies considerably the analysis of the matching of the structure constants across T-duality: firstly notice that
\be
\widehat{\kappa}_{\mathscr{P}} = 2 \, ,
\ee
for all these models, which signals the presence of the two M9 branes. Moreover, since for these models the matching of $\widehat{\kappa}_R$ implies the matching of the corresponding 5d Coulomb branch dimensions, we do not need to list the two invariants separately, in the analysis below. This simplification will be dropped in the study of the more complicated singularities $\mathfrak g = \mathfrak{d}_k$ and $\mathfrak{e}_{6,7,8}$ below. 

\subsection{The $k=2$ examples}
\begin{figure}
\begin{center}
\begin{tabular}{ccc}
\includegraphics[scale=0.4]{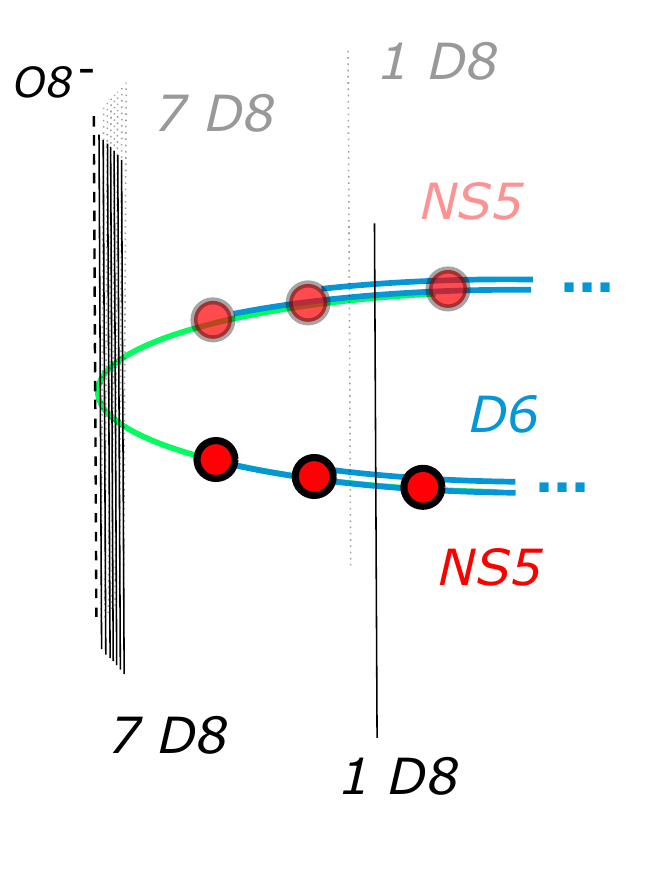}&\includegraphics[scale=0.4]{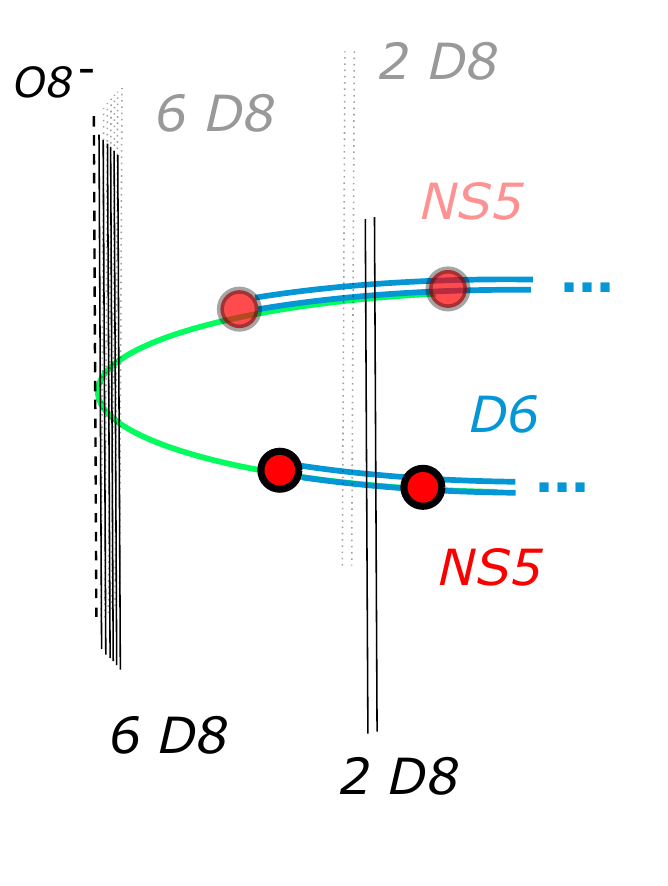}&\includegraphics[scale=0.4]{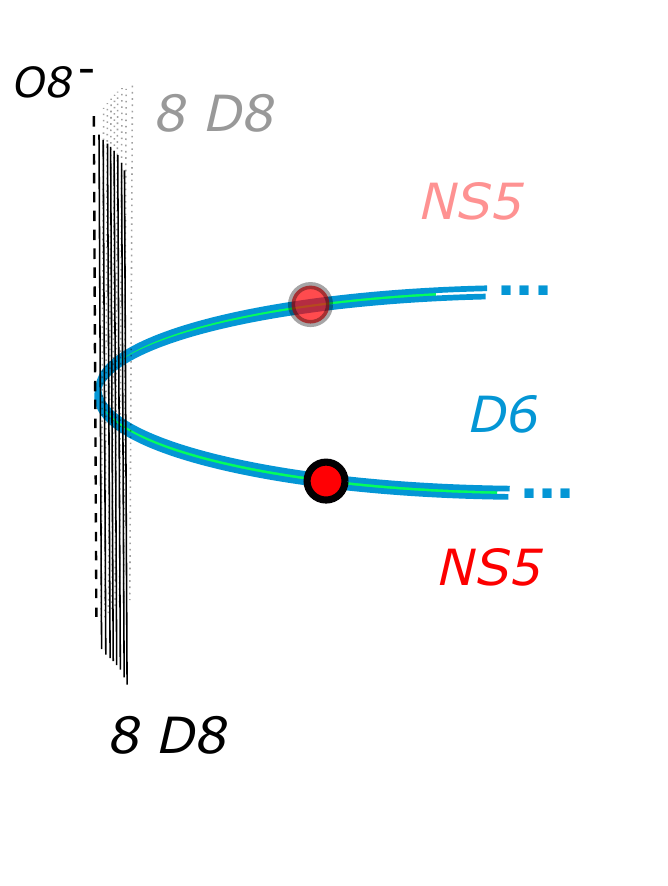}\\
$\boldsymbol{\mu} = 1+1$ & $\boldsymbol{\mu} = 2$ & $\boldsymbol{\mu} = 2'$\\
$N_{\boldsymbol{\mu}} = 2$ \phantom{$\Big|$}& $N_{\boldsymbol{\mu}}  = 1$ & $N_{\boldsymbol{\mu}}  = 0$\\
\end{tabular}
\end{center}
\caption{Type I$'$ brane configurations and corresponding choices of $\boldsymbol{\mu}: \mathbb Z_2 \to E_8$}\label{fig:E8orbinstA1}
\end{figure}

Let us begin with a detailed discussion of the case $k=2$. In this case we have only 3 possible theories of type $\mathcal T(\boldsymbol{\mu},\mathfrak{su}_2)$ corresponding to the following identities of the form \eqref{eq:E8Ak}
\begin{itemize}
\item $2 = 1+1$ with global symmetry $E_8$,
\item $2 = 2$ with global symmetry  $SU(2) \times E_7$,
\item $2 = 2'$ with global symmetry $SO(16)$.
\end{itemize}
The corresponding generalized quivers are
\be\label{eq:TSU2}
\begin{aligned}
&\mathcal T(1+1,\mathfrak{su}_2) = [\fe_8] \, 1 \, 2 \, \underset{[N_f =1]}{\overset{\mathfrak{su}_2}{2}} \, [\fsu_2] \\
&\mathcal T(2,\mathfrak{su}_2) = [\fe_7] \, 1 \, \underset{[N_f =2]}{\overset{\mathfrak{su}_2}{2}} \, [\fsu_2]\\
&\mathcal T(2',\mathfrak{su}_2) = [\fso_{16}] \, \overset{\mathfrak{sp}_1}{1} \, [\fsu_2] \\
\end{aligned}
\ee
which are associated in Type I$'$ to the brane configurations in Figure \ref{fig:E8orbinstA1}.
The conformal matter theory of $N'$ M5 branes along a $\mathbb C^2/\mathbb Z_2$ singularity is
\be
\mathcal T_{N'}(\mathfrak{su}_2,\mathfrak{su}_2) = [\mathfrak{su}_2] \, \underbrace{\overset{\mathfrak{su}_2}{2} \, \overset{\mathfrak{su}_2}{2} \cdots \overset{\mathfrak{su}_2}{2}}_{N'-1} \, [\mathfrak{su}_2]
\ee
Let us consider the fusion operation for the case $\mathcal K_2(1+1,1+1;\mathfrak{su}_2)$: the $SU(2)$ global symmetries of the $\mathcal T_{N'}(\mathfrak{su}_2,\mathfrak{su}_2)$ theories have to be fused into a new gauge node with the $SU(2)$ global symmetries of the two $\mathcal T(1+1,\mathfrak{su}_2)$ theories on the left and on the right as follows:
$$
\begin{aligned}
&\xymatrix@=1.1pc{[\fe_8] \, 1 \, 2 \, \underset{[N_f =1]}{\overset{\mathfrak{su}_2}{2}} \, \textcolor{red}{[\mathfrak{su}_2]}\ar@{-}[r]&\textcolor{red}{[\mathfrak{su}_2] }\, \underbrace{\overset{\mathfrak{su}_2}{2} \, \overset{\mathfrak{su}_2}{2} \cdots \overset{\mathfrak{su}_2}{2}}_{N-3} \, \textcolor{red}{[\mathfrak{su}_2]}&\ar@{red}@{-}[l]\textcolor{red}{[\mathfrak{su}_2]} \, \underset{[N_f =1]}{\overset{\mathfrak{su}_2}{2}}\, 2 \,1 \, [\fe_8]}\\
&\quad\qquad \longrightarrow \qquad   [\fe_8] \, 1 \, 2 \, \underset{[N_f =1]}{\overset{\mathfrak{su}_2}{2}} \, \textcolor{red}{\overset{\mathfrak{su}_2}{2}} \, \underbrace{\overset{\mathfrak{su}_2}{2} \, \overset{\mathfrak{su}_2}{2} \cdots \overset{\mathfrak{su}_2}{2}}_{N-3} \textcolor{red}{\,\overset{\mathfrak{su}_2}{2} }\, \underset{[N_f =1]}{\overset{\mathfrak{su}_2}{2}}\, 2 \,1 \, [\fe_8]
\end{aligned}
$$
Proceeding similarly, we obtain 6 options for the possible instanton LSTs, which we list in Table \ref{tab:E8E8A1all}, where we have highlighted in red the fusion nodes. The two theories $\mathcal K_N(1+1,2';\mathfrak{su}_2)$ and $\mathcal K_N(2,2;\mathfrak{su}_2)$ have the same $\widehat{\kappa}_R$. This gives some evidence that there is a quantum transition which renders the two theories equivalent on a circle, as we shall see below, this is an example where we have a triality at fixed value of $N$.

\medskip

Clearly these results needs to be slightly modified when $N$ is small, for instance we have \cite{DZLO2}
\be
\begin{aligned}
&\mathcal K_2(1+1,1+1; \mathfrak{su}_2) = [\fe_8] \, 1 \, 2 \, \underset{[N_f =1]}{\overset{\mathfrak{su}_2}{2}} \, \textcolor{red}{\,\overset{\mathfrak{su}_2}{2} } \, \underset{[N_f =1]}{\overset{\mathfrak{su}_2}{2}}  \, 2 \, 1 \, [\fe_8] \\
&\mathcal K_1(1+1,1+1; \mathfrak{su}_2) = [\fe_8] \, 1 \,  \,2 \, \textcolor{red}{\underset{[N_f =1]}{\overset{\mathfrak{su}_2}{2}}} \, \textcolor{red}{\underset{[N_f =1]}{\overset{\mathfrak{su}_2}{2}}}  \, 2 \, 1 \, [\fe_8] \
\end{aligned}
\ee
for the cases $N=1,2$ and a choice of $\boldsymbol{\mu}_i$ that is not breaking $E_8$.

\begin{table}
\begin{center}
\begin{tabular}{c|c|c}
$\mathcal K_N(\boldsymbol{\mu}_1,\boldsymbol{\mu}_2;\mathfrak{g})$ \phantom{$\Big|$}& & \phantom{$\Big|$} $\widehat{\kappa}_R$ \\
\hline
&&\\
$\mathcal K_N(2',2'; \mathfrak{su}_2)$ &  $[\fso_{16}] \, \overset{\mathfrak{sp}_1}{1} \, \, \textcolor{red}{\,\overset{\mathfrak{su}_2}{2} }\, \underbrace{\overset{\mathfrak{su}_2}{2} \, \overset{\mathfrak{su}_2}{2} \cdots \overset{\mathfrak{su}_2}{2}}_{N-3} \, \textcolor{red}{\,\overset{\mathfrak{su}_2}{2} } \, \, \overset{\mathfrak{sp}_1}{1} \, [\fso_{16}] $ & $2N + 2$\\
&&\\
$\mathcal K_N(2,2';  \mathfrak{su}_2)$ & $ [\fe_7] \, 1 \, \underset{[N_f =2]}{\overset{\mathfrak{su}_2}{2}} \,  \, \textcolor{red}{\,\overset{\mathfrak{su}_2}{2} }\, \underbrace{\overset{\mathfrak{su}_2}{2} \, \overset{\mathfrak{su}_2}{2} \cdots \overset{\mathfrak{su}_2}{2}}_{N-3} \, \textcolor{red}{\,\overset{\mathfrak{su}_2}{2} } \,\overset{\mathfrak{sp}_1}{1} \, [\fso_{16}]$ & $2N + 3$\\
&&\\
$\mathcal K_N(1+1,2';  \mathfrak{su}_2)$ & $[\fe_8] \, 1 \, 2 \, \underset{[N_f =1]}{\overset{\mathfrak{su}_2}{2}} \,  \, \textcolor{red}{\,\overset{\mathfrak{su}_2}{2} }\, \underbrace{\overset{\mathfrak{su}_2}{2} \, \overset{\mathfrak{su}_2}{2} \cdots \overset{\mathfrak{su}_2}{2}}_{N-3} \, \textcolor{red}{\,\overset{\mathfrak{su}_2}{2} } \, \, \overset{\mathfrak{sp}_1}{1} \, [\fso_{16}]$ & $2N+4$\\
&&\\ 
$\mathcal K_N(2,2;  \mathfrak{su}_2)$ & $ [\fe_7] \, 1 \, \underset{[N_f =2]}{\overset{\mathfrak{su}_2}{2}} \,  \, \textcolor{red}{\,\overset{\mathfrak{su}_2}{2} }\, \underbrace{\overset{\mathfrak{su}_2}{2} \, \overset{\mathfrak{su}_2}{2} \cdots \overset{\mathfrak{su}_2}{2}}_{N-3} \, \textcolor{red}{\,\overset{\mathfrak{su}_2}{2} } \, \, \underset{[N_f =2]}{\overset{\mathfrak{su}_2}{2}} \, 1 \, [\fe_7]$ & $2N+4$ \\
&&\\
$\mathcal K_N(1+1,2; \mathfrak{su}_2)$ &$ [\fe_8] \, 1 \, 2 \, \underset{[N_f =1]}{\overset{\mathfrak{su}_2}{2}} \,  \, \textcolor{red}{\,\overset{\mathfrak{su}_2}{2} }\, \underbrace{\overset{\mathfrak{su}_2}{2} \, \overset{\mathfrak{su}_2}{2} \cdots \overset{\mathfrak{su}_2}{2}}_{N-3} \, \textcolor{red}{\,\overset{\mathfrak{su}_2}{2} } \, \underset{[N_f =2]}{\overset{\mathfrak{su}_2}{2}} \, 1 \, [\fe_7]$& $2N+5$ \\
&&\\
$\mathcal K_N(1+1,1+1; \mathfrak{su}_2)$ & $[\fe_8] \, 1 \, 2 \, \underset{[N_f =1]}{\overset{\mathfrak{su}_2}{2}} \, \textcolor{red}{\,\overset{\mathfrak{su}_2}{2} }\, \underbrace{\overset{\mathfrak{su}_2}{2} \, \overset{\mathfrak{su}_2}{2} \cdots \overset{\mathfrak{su}_2}{2}}_{N-3} \, \textcolor{red}{\,\overset{\mathfrak{su}_2}{2} } \, \underset{[N_f =1]}{\overset{\mathfrak{su}_2}{2}}  \, 2 \, 1 \, [\fe_8]$ & $2N+6$
\end{tabular}
\end{center}
\caption{All the $E_8 \times E_8$ $\mathfrak{a}_1$ instanton LSTs and their corresponding $\widehat{\kappa}_R$.}\label{tab:E8E8A1all}
\end{table}


\medskip

\begin{figure}

\includegraphics[scale=0.6]{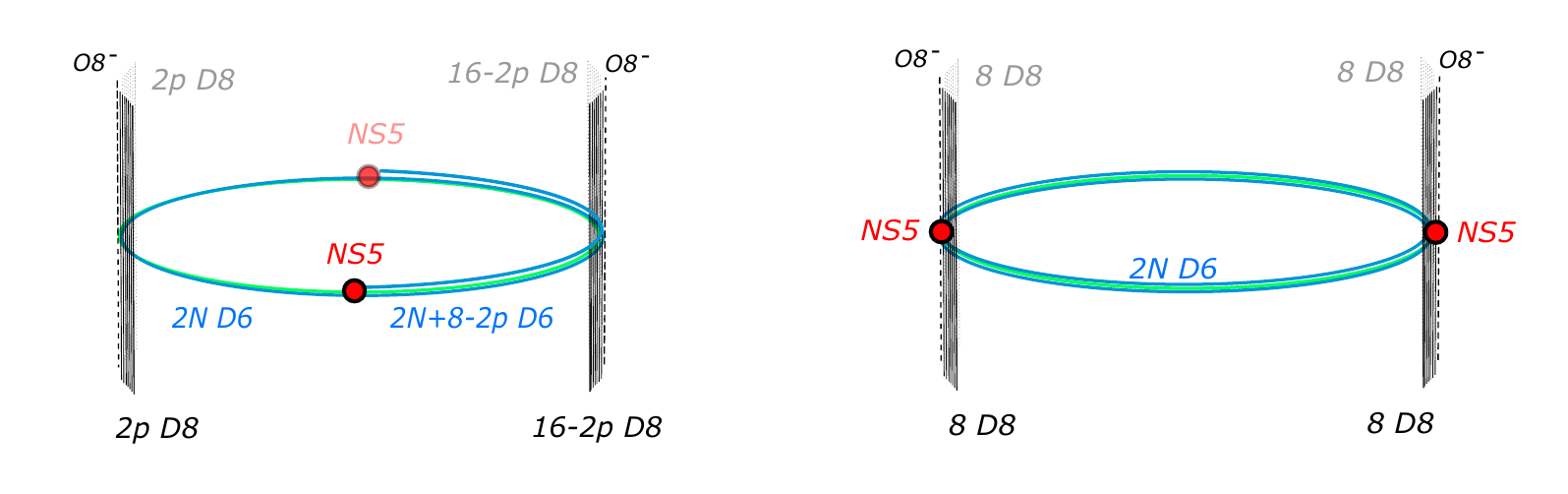}

\begin{center}
\begin{tabular}{c|c|c}
$\widetilde{\mathcal K}_N(\boldsymbol{\lambda};\mathfrak g)$ & & $\widehat{\kappa}_R$\\
\hline
&&\\
$\widetilde{\mathcal K}_N(2p,16-2p;\mathfrak{su}_2)$ & $[\mathfrak{so}_{4p}]\, \overset{\mathfrak{sp}_{N}}{1} \, \overset{\mathfrak{sp}_{N+4-p}}{1} \, [\mathfrak{so}_{32-4p}]$ & $2N + 6 - p$ \\
&&\\
$\widetilde{\mathcal K}_N(8^*,8^*;\mathfrak{su}_2)$ &$ [\mathfrak{u}(8)] \overset{\mathfrak{su}_{2N}}{0} [\mathfrak{u}(8)] $ & $2N$ \\
&&\\
\end{tabular}
\caption{Brane diagrams and generalized quivers for the theories of $Spin(32)/\mathbb Z_2$ instantons on a $\mathbb C^2/\mathbb Z_2$ singularity with $\widetilde{\textsf{w}}_2 = 0$, and corresponding $\widehat{\kappa}_R$.}\label{tab:S32A1all}
\end{center}

\includegraphics[scale=0.6]{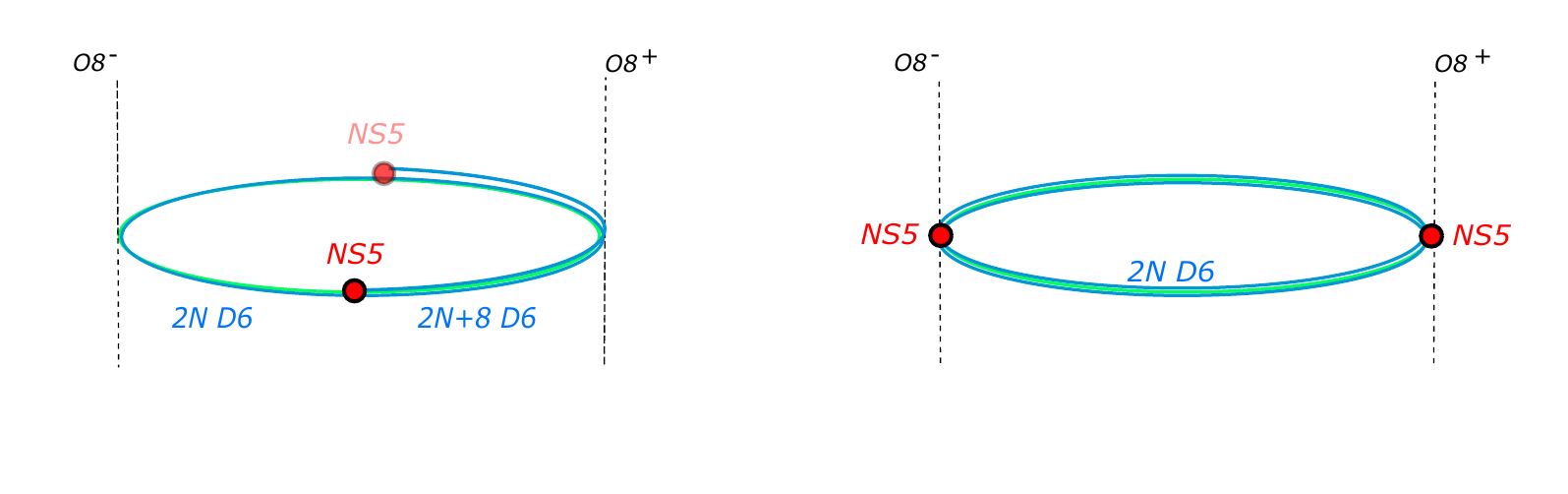}

\caption{$Spin(32)/\mathbb Z_2$ ALE instantons on an $A_1$ singularity -- brane diagrams corresponding to cases without vector structure $\widetilde{\textsf{w}}_2 \neq 0$.}\label{fig:zach}
\end{figure}

Consider now the $Spin(32)/\mathbb Z_2$ side. According to our duality chain we expect to find Type I$'$ systems with $k=2$ NS5 branes and $N$ D6 branes. Naively, following the analysis in \cite{Hanany:1997gh,Brunner:1997gf} we have the possibilities listed in the top part of Figure \ref{fig:zach}, which have also been analyzed as orbifold of the theory of $N$ heterotic $Spin(32)/\mathbb Z_2$ instantons. We point out that, however, one could also obtain models with $\widetilde{\mathsf{w}}_2 \neq 0$, like the ones listed on the bottom part of the figure. In this paper and its sequels \cite{DZLO2,DZLO3} we are going to neglect this possibility, and we will focus on the cases with $\widetilde{\mathsf{w}}_2 =0$.

\medskip

For the cases with $\widetilde{\mathsf{w}}_2 = 0$, we can label $\boldsymbol{\lambda}$ as a splitting $16 = w_1 + w_2$, where $w_i$ are two non-negative even integers. On the brane side this corresponds to the location of the D8 branes with respect to the NS5. We find it conventient to organize such a splitting as 
\be
w_1 = 2p \qquad w_2 = 16 - 2p \qquad 0 \leq p \leq 8
\ee
in order to keep track of the corresponding Romans mass and to account for the brane creation accordingly. There is a further possibility which in the literature is also referred as a case `without vector structure' (abusing slightly terminology\footnote{\ See footnote 2 of \cite{Intriligator:1997kq} for a clarifying discussion.}). This case correspond to the theories in the top right corner of Figure \ref{fig:zach}. Since this case is the unique of its kind for the $\mathbb C^2/\mathbb Z_2$ singularity, we denote the corresponding $\boldsymbol{\lambda}$ by $(w_1,w_2)=(8^*,8^*)$. We describe the resulting generalized quivers in Table \ref{tab:S32A1all}. 

\medskip

Looking at the resulting field theories we see that sometimes a shift in $N$ is necessary for the consistency of the model: for sufficiently small number $N$ of D6 branes and certain values of $p$ one would end up with negative group ranks. The necessity of such shifts is mirrored as the resulting values for $\widehat{\kappa}_R$ are not in the same range as on the T-dual side, which is also further evidence that some shift might be necessary to achieve a matching and there is redundancy among the field theoretical labels for the instantons. This redundancy is the main source of further T-dualities. 

\medskip

Let us discuss the model
\be
\mathcal K_N(1+1,1+1;\mathfrak{su}_2)
\ee
which has
\be
\widehat{\kappa}_R = 2N + 6\,,
\ee
and serves well as an example. We want to compare this model with a T-dual with $\mathfrak{so}_{32}$ flavor symmetry, which is implied by the standard heterotic T-duality. On the T-dual side we see that such a flavor symmetry is achieved by two possible choices of $p$: $p = 8$ or $p=0$, corresponding to the models
\be
\widetilde{K}_{\widetilde{N}}(16,0;\mathfrak{su}_2)\, \quad\text{and}\quad \, \widetilde{K}_{\widetilde{N}}(0,16;\mathfrak{su}_2)\, 
\ee
where we have relabeled the number of NS5 branes on the $Spin(32)/\mathbb Z_2$ side with $\widetilde{N}$ for the sake of comparison. The corresponding generalized quivers are, respectively
\be
[\mathfrak{so}_{32}] \,\, \overset{\mathfrak{sp}_{\widetilde{N}}}{1} \,\, \overset{\mathfrak{sp}_{{\widetilde{N}-4}}}{1} \quad\text{and}\quad \overset{\mathfrak{sp}_{\widetilde{N}}}{1} \,\, \overset{\mathfrak{sp}_{{\widetilde{N}+4}}}{1} \,\, [\mathfrak{so}_{32}]\,.
\ee
The resulting 2-group have structure constants, respectively
\be
\widehat{\kappa}_R = 2 \widetilde{N} - 2 \quad\text{and}\quad \widehat{\kappa}_R = 2 \widetilde{N} + 6
\ee
which suggest the matching for the first model is $\widetilde{N} = N + 4$, hence the desired T-duality is
\be
\widetilde{\mathcal K}_{N+4}(16,0;\mathfrak{su}_2) \sim \mathcal K_N(1+1,1+1;\mathfrak{su}_2)\,.
\ee
For the second model instead, the resulting group ranks from the brane webs are not getting negative for small $N$, hence, the shift is not necessary and indeed we see that we have
\be
\widetilde{\mathcal K}_{N}(0,16;\mathfrak{su}_2) \sim \mathcal K_N(1+1,1+1;\mathfrak{su}_2).
\ee
This latter situation seems to be the case as long as $p\leq 4$, where the gauge theories are automatically consistent for all values of $N$. Assuming this is the case, gives the following T-dualities which are traced by matching the 2-group structure constant $\widehat{\kappa}_R$, keeping $N$ fixed:
\be
\begin{aligned}
&\mathcal K_N(2',2'; \mathfrak{su}_2) \sim \widetilde{\mathcal K}_N(8,8;\mathfrak{su}_2)& \mathcal K_N(2,2'; \mathfrak{su}_2)  \sim \widetilde{\mathcal K}_N(6,10;\mathfrak{su}_2)\\
&\mathcal K_N(1+1,2'; \mathfrak{su}_2) \sim \widetilde{\mathcal K}_N(4,12;\mathfrak{su}_2) &\mathcal K_N(2,2; \mathfrak{su}_2)\sim \widetilde{\mathcal K}_N(4,12;\mathfrak{su}_2)\\
&\mathcal K_N(1+1,2; \mathfrak{su}_2) \sim \widetilde{\mathcal K}_N(2,14;\mathfrak{su}_2) & \mathcal K_N(1+1,1+1; \mathfrak{su}_2) \sim \widetilde{\mathcal K}_N(0,16;\mathfrak{su}_2)
\end{aligned}
\ee

At this point it is natural to ask about the $\widetilde{\mathcal K}_N(8^*,8^*;\mathfrak{su}_2)$ theory. Keeping $N$ fixed, we do not see an immediate T-dual, and for this example there is no need of shifting $N$ for ensuring the positivity of the gauge ranks. However, recall the effect observed in \cite{Aspinwall:1996vc}: in 5d we can trade rank of gauge groups for tensor branch dimensions --- as long as we are ensuring that the corresponding theories have the same 2-group structure constants, Coulomb branch dimensions, and flavor group ranks, the corresponding models are likely to be equivalent in 5d. In particular one is lead to claim the following equivalence
\be
 \widetilde{\mathcal K}_N(8,8;\mathfrak{su}_2) \sim  \widetilde{\mathcal K}_{N+1}(8^*,8^*;\mathfrak{su}_2)
\ee
Including these types of transitions, we see that all these models come in two families: For the Heterotic $Spin(32)/\mathbb Z_2$ we have
\begin{itemize}
\item Family 1: $p = 0 \text{ mod } 2$
$$
\widetilde{\mathcal K}_N(0,16;\mathfrak{su}_2) \sim \widetilde{\mathcal K}_{N+1}(4,12;\mathfrak{su}_2) \sim \widetilde{\mathcal K}_{N+2}(8,8;\mathfrak{su}_2)\sim\widetilde{\mathcal K}_{N+3}(8^*,8^*;\mathfrak{su}_2) \\
$$
\item Family 2: $p = 1 \text{ mod } 2$
$$
\widetilde{\mathcal K}_N(2,14;\mathfrak{su}_2) \sim \widetilde{\mathcal K}_{N+1}(6,10;\mathfrak{su}_2)\\
$$
\end{itemize}
while for the $E_8 \times E_8$ we get
\begin{itemize}
\item Family 1: $\widehat{\kappa}_R = 0 \text{ mod } 2$
$$
\begin{aligned}
\mathcal K_N(1+1,1+1; \mathfrak{su}_2) \sim \mathcal K_{N+1}(2,2; \mathfrak{su}_2) \sim \mathcal K_{N+1}(1+1,2'; \mathfrak{su}_2)\sim \mathcal K_{N+2}(2',2'; \mathfrak{su}_2)\\
\end{aligned}
$$
\item  Family 2: $\widehat{\kappa}_R = 1 \text{ mod } 2$
$$
\mathcal K_N(1+1,2; \mathfrak{su}_2)\sim \mathcal K_{N+1}(2,2'; \mathfrak{su}_2)
$$
\end{itemize}
Then it becomes clear that elements of a fixed Family in either Heterotic models are the ones which can be transformed into one another other upon T-duality.

\medskip

This should not come as a surprise: a similar effect was observed by \cite{Bastian:2017ary,Bastian:2018dfu} in the context of the LSTs of $N$ M5 branes on $\mathbb C^2/\mathbb Z_k$, where all models with the same $\widehat{\kappa}_R = N \times k$ have been proven to be T-dual. Here we are essentially decorating the same geometries with M9 branes, and we are seeing the counterpart of this effect on the Heterotic ALE instanton LSTs. A way to see this effect explicitly is to get to the IIB duality frame in equation \eqref{eq:IIB} and start playing with S-duality and flop transitions for the corresponding $(p,q)$ five-brane webs. The same mechanism already observed in \cite{Bastian:2017ary,Bastian:2018dfu}, adapted to this slightly more general situation, can be used to predict the T-dualities above.

\subsection{Examples: Dualities for LSTs with $SO(16)^2$ symmetry}

\begin{figure}
\begin{center}
\includegraphics[scale=0.5]{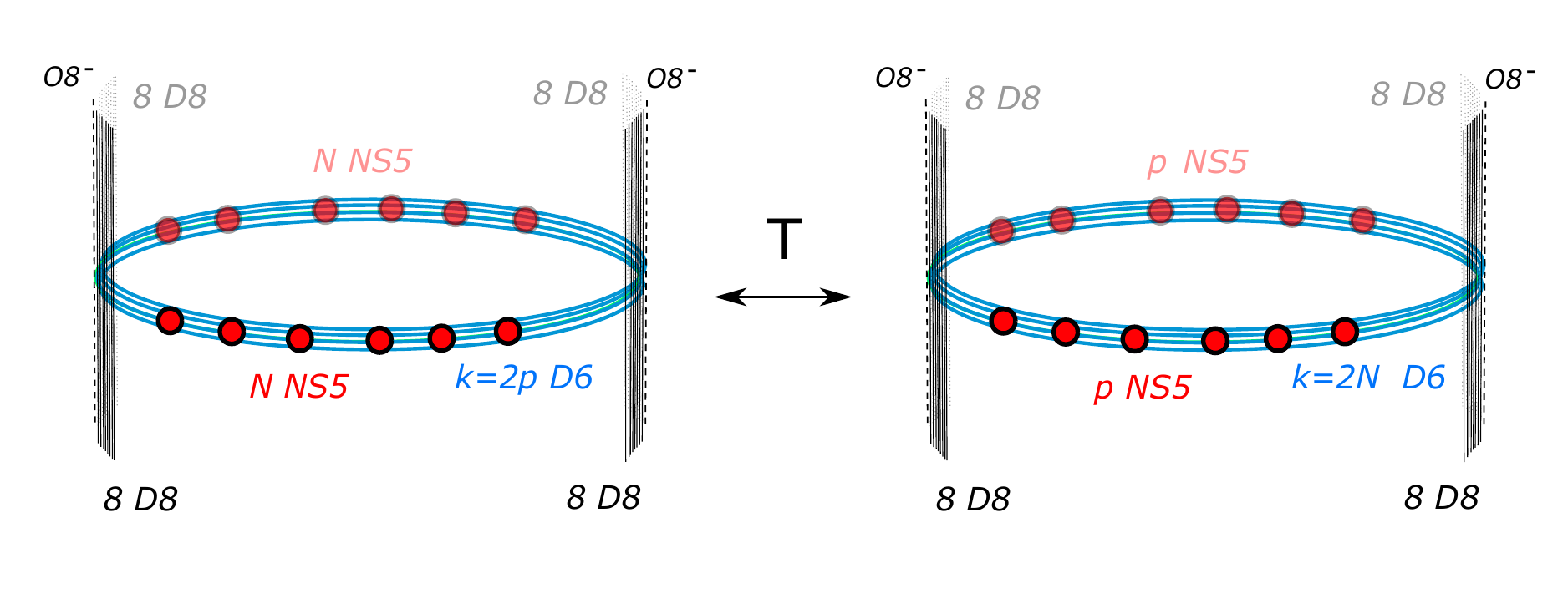}
\end{center}
\caption{\textsc{Left:} Example of Type I$'$ configurations corresponding to $E_8\times E_8$ ALE instanton LSTs of $A$-type. In this figure we have a case $k = 2p$ and $F_a(\boldsymbol{\mu_a}) = SO(16)$ $a=1,2$ corresponding to the decomposition $k = 2 + 2 + 2 + ... + 2$. \textsc{Right:} T-dual configuration corresponding to the $Spin(32)/\mathbb Z_2$ case.}\label{fig:TIpE8}
\end{figure}

In the above examples we have seen that the model $\mathcal K_N(2',2'; \mathfrak g = \mathfrak a_1)$ with tensor branch
\be
[\fso_{16}] \, \overset{\mathfrak{sp}_1}{1} \, \overset{\mathfrak{su}_2}{2} \, \cdots  \, \overset{\mathfrak{su}_2}{2} \,\overset{\mathfrak{sp}_1}{1} \,[\fso_{16}] 
\ee
is T-dual to the model $\widetilde{\mathcal K}_N(8,8; \mathfrak g = \mathfrak a_1,v)$ with tensor branch
\be
[\fso_{16}] \, \overset{\mathfrak{sp}_N}{1} \, \overset{\mathfrak{sp}_N}{1} \, [\fso_{16}] 
\ee
This suggests to seek for a generalization, which is immediate from the duality chain we discussed in the previous section. Indeed, consider the case $k=2p$. If that is the case, we can always decompose
\be
k = 2 p = 2' + 2' + 2' + \cdots + 2' \qquad p \text{ times}
\ee
and hence we expect to be able to construct a 6d LST with 6d global symmetry $SO(16)\times SO(16)$. The latter is realized in Type I$'$ in figure \ref{fig:TIpE8}, and the corresponding generalized quiver is
\be
\mathcal K_N(SO(16),SO(16);\mathfrak g = \mathfrak a_{2p-1}) \, \colon \, [\fso_{16}] \, \overset{\mathfrak{sp}_p}{1} \, \underbrace{\overset{\mathfrak{su}_{2p}}{2} \cdots \overset{\mathfrak{su}_{2p}}{2}}_{N-1} \, \overset{\mathfrak{sp}_p}{1} \, [\fso_{16}]
\ee
Building on the T-duality we discussed above, we obtain that in this case the T-dual model is
\be
[\fso_{16}] \, \underbrace{\overset{\mathfrak{sp}_N}{1} \, \overset{\mathfrak{su}_{2N}}{2} \, \cdots  \, \overset{\mathfrak{su}_{2N}}{2} \,\overset{\mathfrak{sp}_{N}}{1}}_{p+1} \,[\fso_{16}] 
\ee
which gives an interesting (not-simply laced) version of the more familiar fiber base duality:
\be
\mathcal K_N(2',2'; \mathfrak g = \mathfrak a_{2p-1}) \, \leftrightarrow \, \widetilde{\mathcal K}_N(8,0,0,...,0,8; \mathfrak g = \mathfrak a_{2p-1}) 
\ee
Where the theory on the RHS is the theory of $N$ Heterotic $Spin(32)/\mathbb Z_2$  instantons on $\mathbb C^2/\mathbb Z_{2p}$ with a choice of $\boldsymbol{\lambda}$ corresponding to the diagram in Figure \ref{fig:TIpE8} with $p$ and $N$ (base and fiber) swapped.

As we shall see below, this perspective will be useful for determining the corresponding T-duals for this class of examples.

This does not come as a surprise, since it is well known that systems of M5 branes probing singularities have several T-dualities, and here we are essentially promoting them to the situation where we are adding M9 branes.

\subsection{Higher $k$ examples}

One can proceed similarly increasing the value of $k$. In order to have more clear expectations on the generic behavior, we consider in details the cases $k=3$ and $k=4$.

\subsubsection{Heterotic instantons on a $\mathbb C^2/\mathbb Z_3$ singularity}
\begin{figure}
\begin{center}
\begin{tabular}{cc}
\begin{tabular}{c}
\includegraphics[scale=0.4]{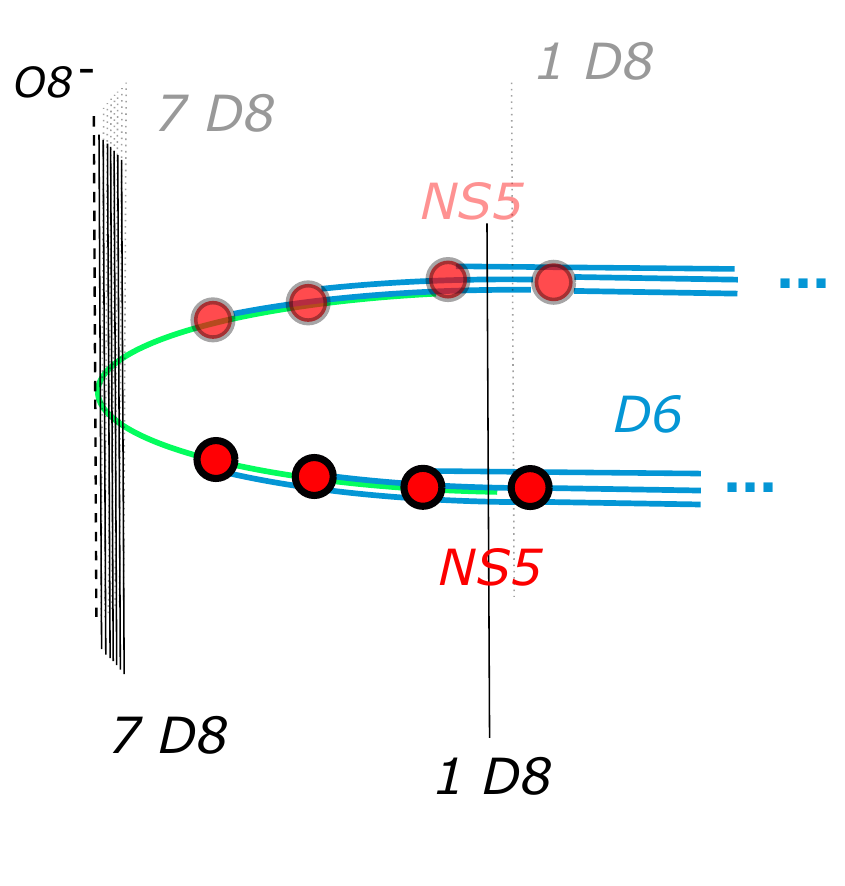} \\
$\boldsymbol{\mu} = 1+1+1$\\
$N_{\boldsymbol{\mu}} = 3$\\
\end{tabular}&\begin{tabular}{|ccc|}
\hline
$\boldsymbol{\mu}$& $\mathcal T(\boldsymbol{\mu},\mathfrak{su}_3)$ \phantom{$\Big|$}& $\widehat{\kappa}_{\mathfrak{su}_3,\boldsymbol{\mu}}$\\
\hline
$\boldsymbol{\mu}=1+1+1$ & $[\fe_8] \, 1 \, 2 \, \overset{\mathfrak{su}_2}{2} \, \underset{[N_f = 1]}{\overset{\mathfrak{su}_3}{2}} [\fsu_3]$ & +7\phantom{$\Big|$}\\
$\boldsymbol{\mu}=2+1$ & $[\fe_7] \, 1 \, \underset{[N_f=1]}{\overset{\mathfrak{su}_2}{2}} \,\, \underset{[N_f =1]}{\overset{\mathfrak{su}_3}{2}} \, [\fsu_3] $& +6\phantom{$\Big|$}\\
$\boldsymbol{\mu} =2'+ 1$ & $[\fso_{14}] \, \overset{\mathfrak{sp}_1}{1} \underset{[N_f = 1]}{\overset{\mathfrak{su}_3}{2}} \, [\fsu_3] \, $&+5\phantom{$\Big|$}\\
$\boldsymbol{\mu}=3$& $[\fe_6] \, 1 \, \underset{[N_f = 3]}{\overset{\mathfrak{su}_3}{2}} \, [\fsu_3]$ & +4\phantom{$\Big|$} \\
$\boldsymbol{\mu}=3'$&$ [\fsu_9] \, \overset{\mathfrak{su}_3}{1} \, [\fsu_3]$&+3\phantom{$\Big|$}\\
\hline
\end{tabular}
\\
\includegraphics[scale=0.4]{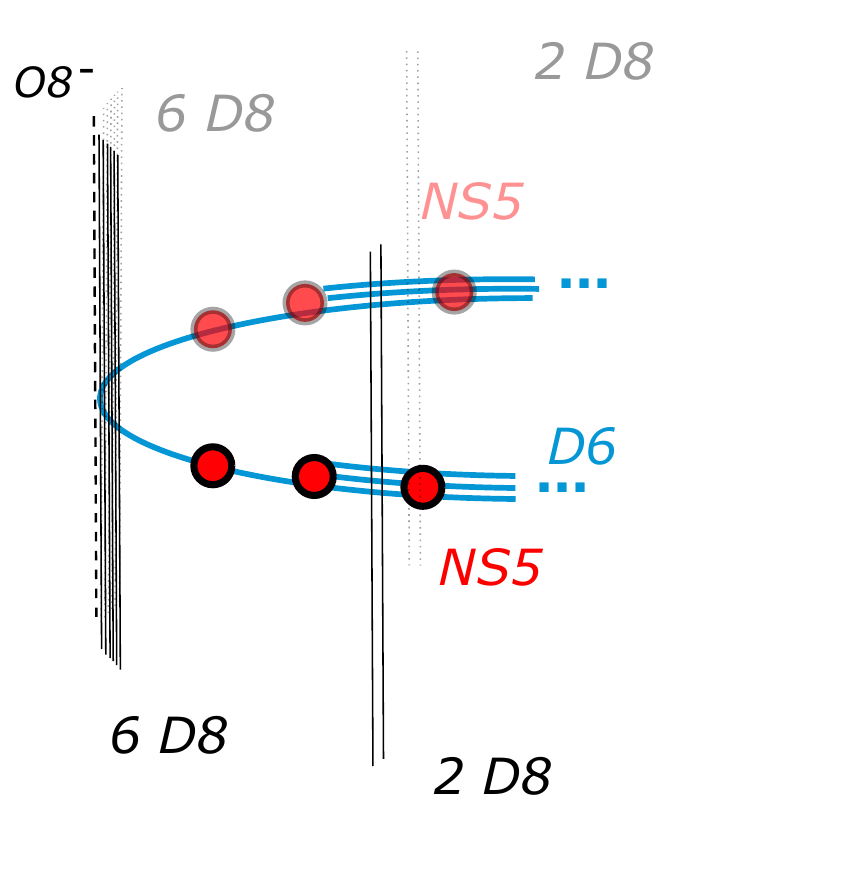}  &\includegraphics[scale=0.4]{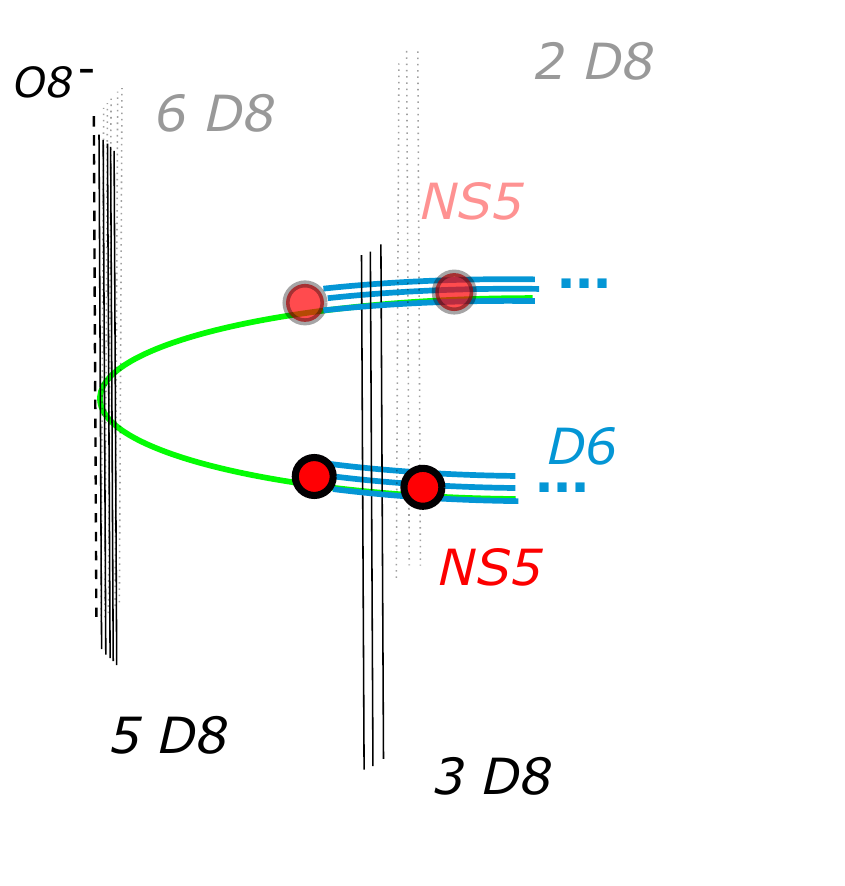}\\
$\boldsymbol{\mu} = 2+1$& $\boldsymbol{\mu} =3$\\
$N_{\boldsymbol{\mu}} = 2$&$N_{\boldsymbol{\mu}} = 1$\\
 \includegraphics[scale=0.4]{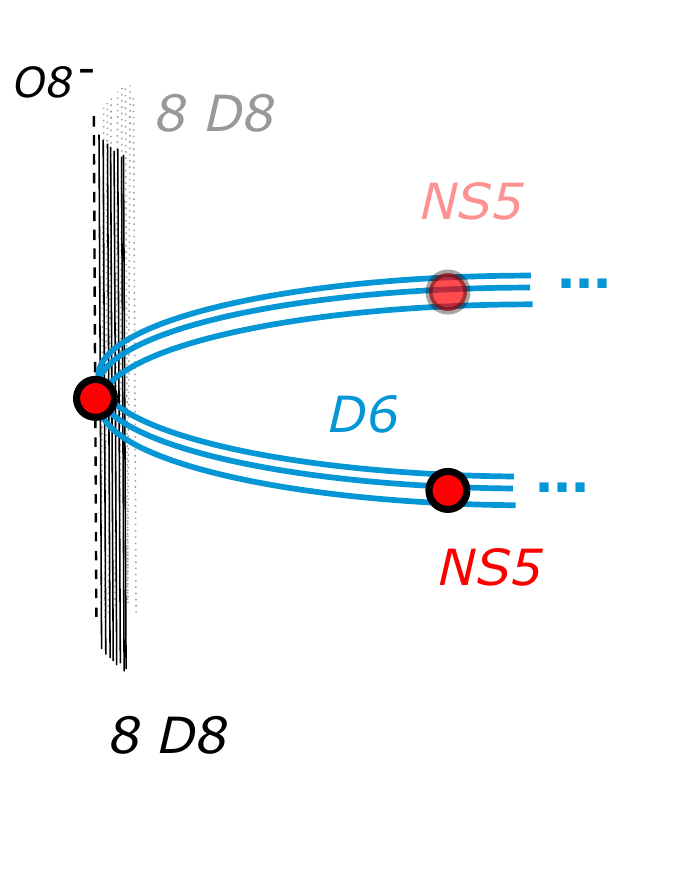}&\includegraphics[scale=0.4]{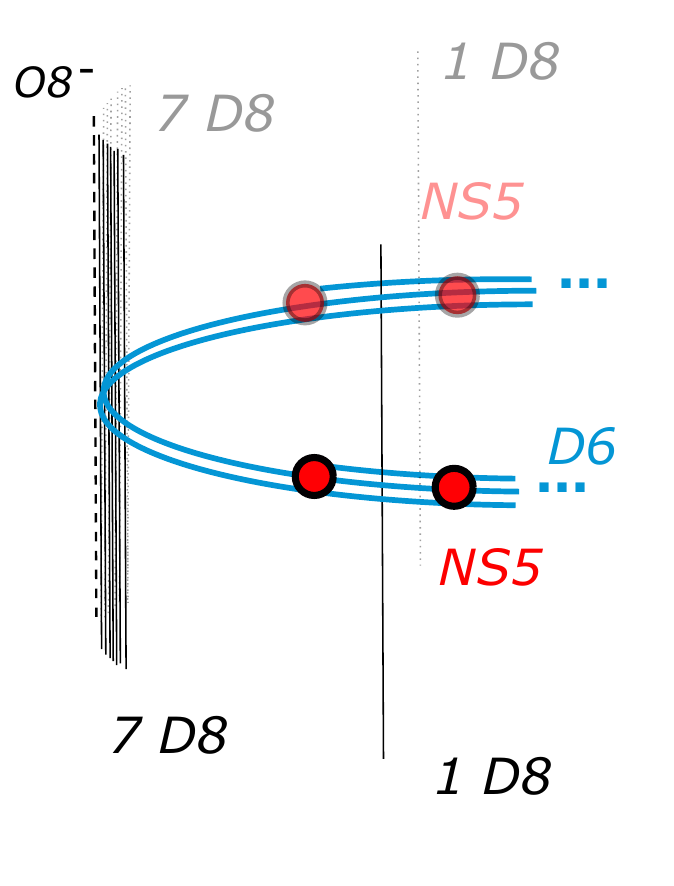}\\
 $\boldsymbol{\mu} = 3'$&$\boldsymbol{\mu} = 2'+1$\\
$N_{\boldsymbol{\mu}} = 0$ &$N_{\boldsymbol{\mu}} = 1$\\
\end{tabular}
\caption{The $\mathcal T(\boldsymbol{\mu},\mathfrak{su}_3)$ SCFTs corresponding to the embeddings $\boldsymbol{\mu}: \mathbb Z_3 \to E_8$ and the dual configurations in Type I$'$. We indicate for later convenience also the number $N_{\boldsymbol{\mu}}$ of additional NS5 branes which arises from M9 fractionalization.}\label{fig:vitadimerda}
\end{center}
\end{figure}

For $k=3$ we have a total of 5 different theories of type $\mathcal T(\boldsymbol{\mu},\mathfrak{su}_3)$, which we list in Figure \ref{fig:vitadimerda}. All these models have $\ell =1$. The corresponding brane configurations in Type I$'$ are listed in Figure \ref{fig:vitadimerda}.  The theory corresponding to the LST of $N+1$ heterotic instantons with prescribed flat connections at infinity can be easily assembled from these data. For sufficiently large $N$, the desired theory has the form
\be
\xymatrix{\mathcal T(\boldsymbol{\mu_1},\mathfrak{su}_3) \ar@{-}[r]^{\mathfrak {su}_3} &\mathcal T_{N-2}(\mathfrak{su}_3,\mathfrak{su}_3)  \ar@{-}[r]^{\mathfrak {su}_3}&\mathcal T(\boldsymbol{\mu_2},\mathfrak{su}_3) }
\ee
in order to determine the 2-group structure constant for the corresponding theory one just has to read off $\mathcal T(\boldsymbol{\mu_a},\mathfrak{su}_3)$  and perform a fission on the common diagonal with the conformal matter theory of $N-2$ M5 branes along an $\mathbb C^2/\mathbb Z_3$ singularity:
\be
\mathcal T_{N-2}(\mathfrak{su}_3,\mathfrak{su}_3) = [\fsu_3] \, \underbrace{\overset{\mathfrak{su}_3}{2} \, \overset{\mathfrak{su}_3}{2} \cdots \overset{\mathfrak{su}_3}{2}}_{N-3} \, [\fsu_3].
\ee
The latter contributes to $\widehat{\kappa}_R$ with
\be
3(N-3)
\ee
that, together with the two extra contributions from the fusion nodes gives
\be
\mathcal K_N(\boldsymbol{\mu}_1,\boldsymbol{\mu}_2;\mathfrak{su}_3) \quad\text{ has }\quad \widehat{\kappa}_R = 3N - 3 + \widehat{\kappa}_{\mathfrak{su}_3,\boldsymbol{\mu}_1} + \widehat{\kappa}_{\mathfrak{su}_3,\boldsymbol{\mu}_2}\,.
\ee

As a concrete example, consider the case $\boldsymbol{\mu}_1=2'+1$ while $\boldsymbol{\mu}_2=2+1$:
\be
\mathcal K_N(2'+1,2+1; \mathfrak{su}_3) = [\fe_7] \, 1 \, \underset{[N_f=1]}{\overset{\mathfrak{su}_2}{2}} \,\, \underset{[N_f =1]}{\overset{\mathfrak{su}_3}{2}} \, \textcolor{red}{\overset{\mathfrak{su}_3}{2}} \,\underbrace{ \overset{\mathfrak{su}_3}{2} \, \cdots \overset{\mathfrak{su}_3}{2}}_{N-3}\,  \textcolor{red}{\overset{\mathfrak{su}_3}{2}} \,  \underset{[N_f = 1]}{\overset{\mathfrak{su}_3}{2}}  \, \overset{\mathfrak{sp}_1}{1} \, [\fso_{14}]
\ee
where we are indicating in red the fission gauge nodes and 
\be
\widehat{\kappa}_R = 3N - 3 + 6 + 5 = 3N + 8
\ee
Using the data in Figure \ref{fig:vitadimerda} one can clearly see that we obtain a total of 15 sequences of theories, and that, by shifting $N$, we expect these models are organized in 3 families, corresponding to the values of $\widehat{\kappa}_R$ modulo 3. 

\begin{figure}
\begin{center}
\includegraphics[scale=0.5]{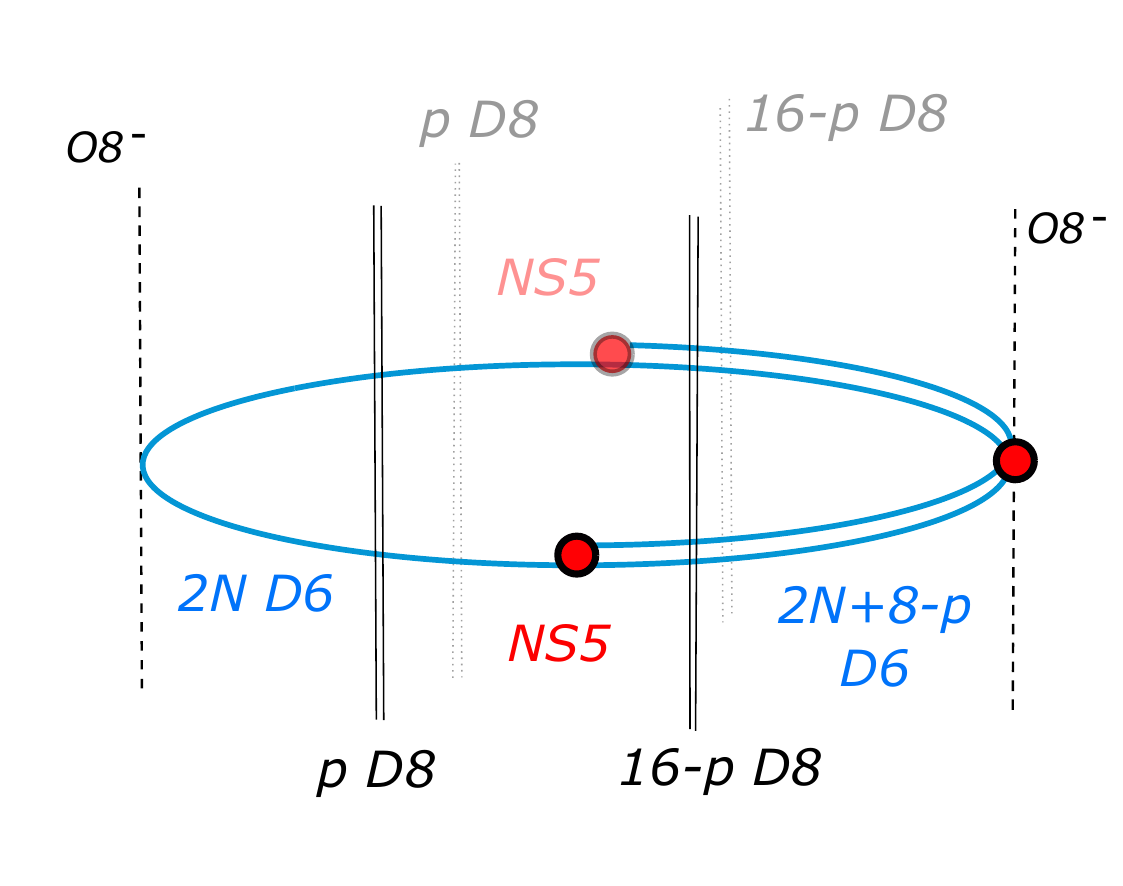}
\end{center}
\caption{Brane diagram for the $\widetilde{\mathcal K}_N(w_1,w_2;\mathfrak{su}_3)$ LSTs with $w_1 = p, w_2 = 16-p$.}\label{fig:Spin32onA2}
\end{figure}

The dual theories in this class are obtained straightforwardly. One has a Type I$'$ configuration with 3 NS5 branes, of which one must be stuck at the O8$^-$ plane -- see Figure \ref{fig:Spin32onA2}. In total we obtain naively a series of 17 models:
\be
\widetilde{\mathcal K}_N(p,16-p;\mathfrak{su}_3) = [\mathfrak{so}_{2p}] \, \overset{\mathfrak{sp}_N}{1} \, \, \underset{[N_A=1]}{\overset{\mathfrak{su}_{2N+8 - p}}{1}}\, \, [\mathfrak{u}_{16-p}] \qquad\qquad 0 \leq p \leq 16\,,
\ee
which have the following structure constants
\be
\widehat{\kappa}_R =  3N+ 9 - p \,.\ee
Similar to the case $k=2$, in order to match one class of models to the other it is necessary to match the corresponding families modulo 3 and sometimes in order to achieve a matching it is necessary to shift $N$, which is also clear from the fact that for small $N$ and $p$ high enough one might end up with negative group ranks. Again as an example we consider the model
\be
\mathcal K_N(1+1+1,1+1+1;\mathfrak{su}_3)
\ee
which has $E_8 \times E_8$ flavor symmetry. We expect this theory is T-dual to the case $p=16$ because of the usual matching of flavor symmetries which underlies the heterotic T-duality, we desire a dual theory with $\mathfrak{so}(32)$ flavor symmetry. The two theories have 
\be
\widehat{\kappa}_R = 3N + 11 \qquad \widehat{\kappa}_R = 3 \widetilde{N} - 7
\ee
where we denoted with $\widetilde{N}$ the number of NS5 branes on the T-dual side for the sake of comparision. This suggests that 
\be\label{eq:bwahahaha}
\widetilde{N} = N + 6
\ee
on the $Spin(32)/\mathbb Z_2$ side. Using this model to normalize $N$ with respect to $\widetilde{N}$ we see that the resulting $\widehat{\kappa}_R$ indeed always match.

\subsubsection{A qualitative explanation from an inequivalent duality chain}

One might wonder whether we could track the origin of the shifts in $N$ which are necessary from the brane web duality chains. Perhaps the most explicit way to do this is to revisit the duality chain we discussed in section \ref{sec:duality1}. We know that in Type $I'$ the heterotic instanton $E_8 \times E_8$ LSTs on a $\mathbb C^2/\mathbb Z_k$ are realized with a slightly larger number of branes $M$, corresponding to the fractionalization of the M9 brane. Let us consider dualizing the diagram in equation \eqref{eq:IIAbrain} along $x^5$, we have
\be\label{eq:IIBbrain-bis}
\begin{tabular}{c|ccccccccccc}
&0 & 1 & 2 & 3 & 4 & 5$'$ & 6 & 7 & 8  & 10\\
\hline
O7-D7 & $\bullet$ & $\bullet$ & $\bullet$ & $\bullet$ & $\bullet$&  & $\bullet$& $\bullet$& $\bullet$&  & \\ 
$M$ NS5 & $\bullet$ & $\bullet$ & $\bullet$ & $\bullet$ & $\bullet$ & $\bullet$ \\
$k$ D5 & $\bullet$ & $\bullet$ & $\bullet$ & $\bullet$ & $\bullet$ &  & & & &  $\bullet$ \\
\end{tabular}
\ee
where now $M=N+N_{\boldsymbol{\mu}_1} + N_{\boldsymbol{\mu}_2}$. Now performing IIB S-duality is swapping the D5 and the NS branes in this picture, and we obtain and equivalent dual picture
\be\label{eq:IIBbrain-bis-b}
\begin{tabular}{c|ccccccccccc}
&0 & 1 & 2 & 3 & 4 & 5$'$ & 6 & 7 & 8  & 10\\
\hline
O7-D7 & $\bullet$ & $\bullet$ & $\bullet$ & $\bullet$ & $\bullet$&  & $\bullet$& $\bullet$& $\bullet$&  & \\ 
$M$ D5 & $\bullet$ & $\bullet$ & $\bullet$ & $\bullet$ & $\bullet$ & $\bullet$ \\
$k$ NS5 & $\bullet$ & $\bullet$ & $\bullet$ & $\bullet$ & $\bullet$ &  & & & &  $\bullet$ \\
\end{tabular}
\ee
at this point we can T-dualise back to IIA using the 10-th direction, giving
\be\label{eq:IIAbrain-bis}
\begin{tabular}{c|ccccccccccc}
&0 & 1 & 2 & 3 & 4 & 5$'$ & 6 & 7 & 8  & 10$'$\\
\hline
O8-D8 & $\bullet$ & $\bullet$ & $\bullet$ & $\bullet$ & $\bullet$&  & $\bullet$& $\bullet$& $\bullet$&  $\bullet$\\ 
$M$ D6 & $\bullet$ & $\bullet$ & $\bullet$ & $\bullet$ & $\bullet$ & $\bullet$ &&&&$\bullet$ \\
$k$ NS5 & $\bullet$ & $\bullet$ & $\bullet$ & $\bullet$ & $\bullet$ &  & & & &  $\bullet$ \\
\end{tabular}
\ee
which indeed coincides to the Type I$'$ description of a system of $M$ heterotic $Spin(32)/\mathbb Z_2$ instantons probing a $\mathbb C^2/\mathbb Z_k$ singularity. This different duality chain, gives a qualitative explanation of the shifts between $N$ and $\widetilde{N}$ we have observed above. Indeed, from the tensor branch of the model $\mathcal T(\boldsymbol{\mu},\mathfrak{su}_3)$ and its dual brane realization in Type I$'$ in Figure \ref{fig:vitadimerda} we read off that $N_{\boldsymbol{\mu}=1+1+1}=3$, and hence for the $E_8 \times E_8$ model with $\boldsymbol{\mu}_1= \boldsymbol{\mu}_2 = \{ 1+1+1 \}$ that we discussed above, we have $M = N + 3 + 3$, consistently with our remark in equation \eqref{eq:bwahahaha} in the previous section.

\medskip

This further duality chain indicates that (as expected) T-dualities often involve models which are related also by shifts in the instanton number $N$, as we remarked above.

\subsubsection{The case of a $\mathbb C^2 / \mathbb Z_4$ singularities}

\begin{table}
\begin{center}
\begin{tabular}{l|l||c|c}
$\boldsymbol{\mu}$ & $F(\boldsymbol{\mu})$ & $\mathcal T(\boldsymbol{\mu},\mathfrak{su}_4)$ \phantom{\Big|} & $\widehat{\kappa}_{\boldsymbol{\mu,\mathfrak{su}_4}}$ \\
\hline
&&\\
$1+1+1+1$ & $E_8$ & $[\fe_8] \, 1 \, 2 \, \overset{\mathfrak{su}_2}{2} \,  \overset{\mathfrak{su}_3}{2} \, \underset{[N_f =1]}{\overset{\mathfrak{su}_4}{2}} \, [\fsu_4]$ & 11  \\ 
&&\\
\hline
&&\\
$1+1+2$  & $E_7$ & $[\fe_7] \, 1 \, \underset{[N_f =1]}{\overset{\mathfrak{su}_2}{2}} \,  \overset{\mathfrak{su}_3}{2}  \, \underset{[N_f =1]}{\overset{\mathfrak{su}_4}{2}} \,  [\fsu_4]$ &10 \\
&&\\
\hline
&&\\
$1+1+2'$ & $SO(14)$ &  $[\fso_{14}] \, \overset{\mathfrak{sp}_1}{1} \, \overset{\mathfrak{su}_3}{2} \, \underset{[N_f =1]}{\overset{\mathfrak{su}_4}{2}} \, [\fsu_4]$ & 9 \\
&&\\
\hline
&&\\
$1+3$ & $E_6 \times SU(2)$ &  $[\fe_6] \, 1 \,\underset{[\fsu_2]}{\overset{\mathfrak{su}_3}{2}} \, \underset{[N_f =1]}{\overset{\mathfrak{su}_4}{2}} \,  [\fsu_4]$ \phantom{\Big|} & 8 \\
&&\\
\hline
&&\\
$1+3'$ & $SU(8)$ & $[\fsu_8] \, \overset{\mathfrak{su}_3}{1} \, \underset{[N_f = 1]}{\overset{\mathfrak{su}_4}{2}} \, [\fsu_4] $ & 7\\
&&\\
\hline
&&\\
$2+2$ & $SU(2) \times E_7$ & $[\fe_7] \, 1 \, \overset{\mathfrak{su}_2}{2} \, \underset{[\fsu_2]}{\overset{\mathfrak{su}_4}{2}} \, [\fsu_4] $ & 7 \\
&&\\
\hline
&&\\
$2+2' $& $SU(2) \times SO(12)$ & $[\fso_{12}] \, \overset{\mathfrak{sp}_1}{1} \, \underset{[\fsu_2]}{\overset{\mathfrak{su}_4}{2}} \, [\fsu_4] $ & 6\\
&&\\
\hline
&&\\
 $ 2'+2' $& $SO(16)$ & $[\fso_{16}] \, \overset{\mathfrak{sp}_2}{1} \, \underset{[\fsu_2]}{\overset{\mathfrak{su}_4}{2}} \, [\fsu_4] $ & 6\\\
 &&\\
\hline
&&\\
 $4$ & $SU(4) \times SO(10)$ & $[\fso_{10}] \, 1 \, \underset{[\fsu_4]}{\overset{\mathfrak{su}_4}{2}} \, [\fsu_4] $ & 5\\
 &&\\
\hline
&&\\
 $4'$ & $SU(2) \times SU(8)$ & $[\fsu_8] \, \underset{[N_A = 1]}{\overset{\mathfrak{su}_4}{1}} \, [\fsu_4] $ & 4\\
 &&\\
\end{tabular}
\end{center}
\caption{$\boldsymbol{\mu}$ VS $F$ for the case $k=4$ \cite{Mekareeya:2017jgc}.}\label{tab:AkexF}
\end{table}

\begin{figure}
\begin{center}
\includegraphics[scale=0.45]{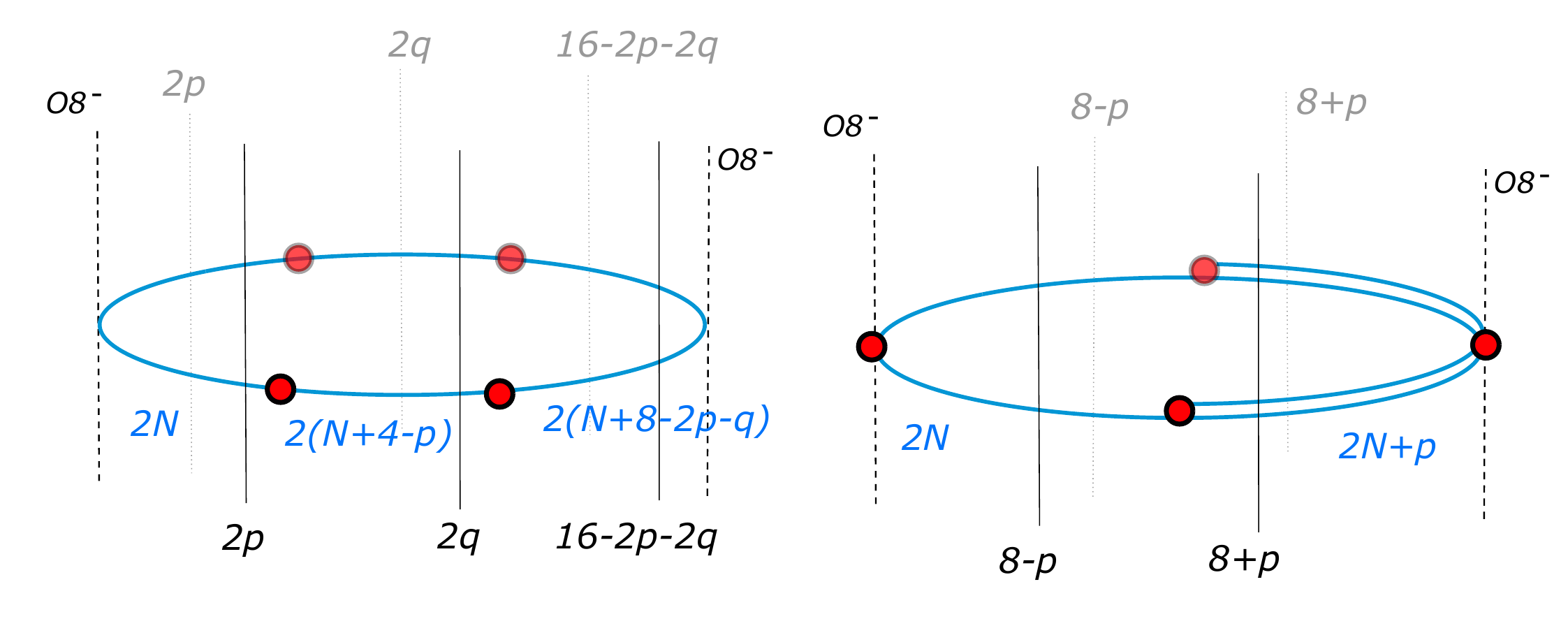}
\caption{Type I$'$ configurations for $Spin(32)/\mathbb{Z}_2$ instantons along a $\mathbb C^4/\mathbb Z_4$ singularity.}\label{fig:s32A3}
\end{center}
\end{figure}

As a further example, in Table \ref{tab:AkexF}, all the theories $\mathcal T(\boldsymbol{\mu},\mathfrak{su}_4)$ can be found. One can proceed as above, \textit{mutatis mutandis}. The theories $\mathcal K_N(\boldsymbol{\mu}_1,\boldsymbol{\mu}_2,\mathfrak{su}_4)$ have
\be
\widehat{\kappa}_R = 4(N - 1) + \widehat{\kappa}_{\boldsymbol{\mu}_1,\mathfrak{su}_4} + \widehat{\kappa}_{\boldsymbol{\mu}_2,\mathfrak{su}_4}
\ee
where $4(N - 1)$ is the contribution from the $\mathfrak{su}_4$ conformal matter and the fusion nodes. We obtain the following values of 
\be
\widehat{\kappa}_R = 4N - 4 + p \qquad\qquad 8 \leq p \leq 22
\ee
As an example consider theory with $\boldsymbol{\mu}_1 = \{1+ 1 + 1 + 1\}$ and $\boldsymbol{\mu}_2 = \{1+3\}$: by fusion we obtain the theory of $N$ heterotic instantons corresponding to these two choices is given by
\be
\mathcal K_N(1+1+1+1;1+3; \mathfrak{su}_4) = [\fe_8] \, 1 \, 2 \, \overset{\mathfrak{su}_2}{2} \,  \overset{\mathfrak{su}_3}{2} \,\underset{[N_f=1]}{\overset{\mathfrak{su}_4}{2}} \, \textcolor{red}{\overset{\mathfrak{su}_4}{2}}\, \underbrace{\overset{\mathfrak{su}_4}{2} \, \overset{\mathfrak{su}_4}{2} \cdots \overset{\mathfrak{su}_4}{2}}_{N-3} \,\textcolor{red}{\overset{\mathfrak{su}_4}{2}}\,  \underset{[N_f=2]}{\overset{\mathfrak{su}_4}{2}}  \,\underset{[\fsu_2]}{\overset{\mathfrak{su}_3}{2}} \, 1 \, [\fe_6]
\ee
where we are indicating in red the fission gauge nodes.

\medskip

The dual models in $Spin(32)/\mathbb Z_2$ are described via Type I$'$ brane diagrams in Figure \ref{fig:s32A3}. We obtain theories of two kinds for models with $\textsf{w}_2 = 0$, namely
\be
\widetilde{\mathcal K}_N(2p,2q,16-2p-2q; \mathfrak{su}_4)= [\mathfrak{so}_{4p}]  \, \overset{\mathfrak{sp}_{N+8}}{1} \,\, \underset{[N_f=2q]}{\overset{\mathfrak{su}_{2(N + 4 - p)}}{2}} \, \, \overset{\mathfrak{sp}_{N+8 - 2p - q}}{1} \, [\mathfrak{so}_{32 - 4p - 4 q}]
\ee
which is defined for pairs $0 \leq p,q \leq 8$ such that $p+q = 8$ give rise to
\be
\widehat{\kappa}_R = 4N + 18 - 4p - q 
\ee
as well as
\be
\widetilde{\mathcal K}_N(8^*-p,8^*+p; \mathfrak{su}_4)= [\mathfrak{u}_{p}]  \, \underset{[N_A=1]}{\overset{\mathfrak{su}_{2N}}{1} }\,\,\underset{[N_A=1]}{ \overset{\mathfrak{su}_{2N + p}}{1}} \, [\mathfrak{u}_{16-p}]
\ee
which is defined for $0\leq p \leq 16$ and gives
\be
\widehat{\kappa}_R = 4N + p
\ee
Also here we observe that only the value of $\widehat{\kappa}_R$ modulo 4 is relevant to establish boundaries for the families of T-dual models.

\subsubsection{Generic behaviour for Heterotic $\mathbb C^2/\mathbb Z_k$ instantons}

Out of these examples we expect the following generic behaviour for Heterotic $\mathbb C^2/\mathbb Z_k$ instantons:
\begin{itemize}
\item The field theoretical labels $\boldsymbol{\lambda}: \mathbb Z_k \to Spin(32)/\mathbb Z_2$ and $\boldsymbol{\mu}_a : \mathbb Z_k \to E_8$ $a=1,2$ are oftentimes confused upon T-duality, moreover multiple T-dual channels open up once shifts in $N$ are allowed corresponding to the fractionalization of M9 branes;
\item There are however always $k$ inequivalent classes of models that do not transition one another upon T-duality, which are distinguished by the value of
\be\label{eq:moddds}
\widehat{\kappa}_R \text{ mod } k
\ee
We expect that models in the same class end up being T-dual upon allowing shifting in $N$.
\item For this class of examples the rank of the flavor symmetry never jumps, moreover the Coulomb branch dimensions are also very closely related to the actual value of $\widehat{\kappa}_R$ as remarked in \cite{DelZotto:2020sop}. Models corresponding to singularities of $\mathfrak d_k$ and $\mathfrak{e}_{6,7,8}$ types exhibit a qualitatively different behaviour in this respect.
\end{itemize}

\section{D-type cases: adding orientifolds to the webs}\label{sec:duality3}

When the singularities involved are of type $D_k$, the corresponding Lie algebra associated by the MacKay correspondence is $\mathfrak g=\mathfrak{so}_{2k}$, obtained from the binary dihedral finite subgroup of $SU(2)$. The presence of such a singularity has a relatively simple effect with respect to the Type I$'$ brane diagrams we have considered in the previous section: it amounts to adding an orientifold six-plane along the locus of the D6 branes. The orientifold six plane changes sign when crossing the NS5 branes, which modifies its contribution to the cosmological constant involved in the various processes of brane creation. Correspondingly one expects to obtain orthosymplectic generalized quivers from this setup. There is one major caveat, however: in all these models the O6 planes are intersecting the O8 planes located at the two ends of the $S^1/\mathbb Z_2$ direction. There are some further degrees of freedom that are trapped at the intersection, which give rise to the ON$^0$ plane of \cite{Hanany:1999sj}. In presence of stuck D8 branes, the corresponding dynamics becomes more interesting.

\medskip

Let us begin from the $Spin(32)/\mathbb Z_2$ cases. In this context, whenever the order $k$ of the $D_k$ singularity is even, one can consider instantons with a non-trivial $\widetilde{\textsf{w}}_2$, i.e. instantons which obstructs a vector structure. Again, these are mapped to LST configurations corresponding to systems of $O8^-$ and $O8^+$, which require dual F-theory configurations which are of the frozen kind. We plan to return to these configurations in future work. For configurations with $\widetilde{\textsf{w}}_2 = 0$, instead, we have configurations with two $O8^-$ located at the two extrema of the interval $S^1/\mathbb Z_2$ with a total of 16 D8s suspended in between. The theory of $N$ instantons corresponds to the presence of $2N$ D6s, located along an O6 plane. The $D_k$ singularity is realized by $2k$ NS5 branes with two ON$^0$ planes located at the intersection of the O6 planes with the O8$^-$ planes. There is a different qualitative behaviour depending on whether $k$ is even or odd, corresponding to the fact that in one case the O8$^-$ plane intersects an O6$^-$ while in the other it ends up intersecting an O6$^+$ plane. The resulting quivers have the same structure already determined by Intriligator and Blum. In this context we cannot have systems with stuck NS5 branes along the O8$^-$ planes because the change in sign of the O6$^-$ across such a stuck NS5 would break the $\mathbb Z_2$ symmetry necessary for the orbifold.

\medskip

For brevity, here we just describe some examples corresponding to the $D_4$ singularity. On the $Spin(32)/\mathbb Z_2$ side, the latter gives rise to instantonic configurations of the form
\be
\begin{matrix}
\underset{[\mathfrak{so}(2w_1)]}{\overset{\mathfrak{sp}_{v_1}}{1}}&&\underset{[\mathfrak{so}(2w_3)]}{\overset{\mathfrak{sp}_{v_3}}{1}}\\
&\underset{[\mathfrak{sp}(w_2)]}{\overset{\mathfrak{so}_{2v_2}}{4}}&\\
\underset{[\mathfrak{so}(2w_4)]}{\overset{\mathfrak{sp}_{v_4}}{1}}&&\underset{[\mathfrak{so}(2w_5)]}{\overset{\mathfrak{sp}_{v_5}}{1}}\\
\end{matrix}
\ee
with LS charge $(1,1,1,1,1)$. The form of the 2-group structure constant for these configurations is
\be
\widehat{\kappa}_R = 2 + v_1 + 2 v_2 + v_3 + v_4 + v_5
\ee
but of course only a few of the $v_i$'s above satisfy anomaly cancellation. An example that will be useful below is provided by the following configuration\footnote{\ Which is realized by the brane diagram in Figure 21 of \cite{Hanany:1999sj}.}
\be
\begin{matrix}
\overset{\mathfrak{sp}_{N}}{1}&&\overset{\mathfrak{sp}_{N}}{1}\\
&\overset{\mathfrak{so}_{4N+16}}{4}&\\
\overset{\mathfrak{sp}_{N}}{1}&&\underset{[\mathfrak{so}(32)]}{\overset{\mathfrak{sp}_{N+8}}{1}}\\
\end{matrix}\qquad \qquad \widehat{\kappa}_R = 8N + 26
\ee

\medskip



Let us consider the $E_8\times E_8$ case realized as $N$ $M5$ branes in the Hořava-Witten setup transverse to a D-type singularity $\mathbb C^2 / \mathbb D_k$ (see Figure \ref{fig:HWframe}). To trace the duality to Type I$'$ we summarize the position of the relevant branes and singularities as follows
\be
\begin{tabular}{c|ccccccccccc}
&0 & 1 & 2 & 3 & 4 & 5 & 6 & 7 & 8 & 9 & 10 \\
\hline
M9 & $\bullet$ & $\bullet$ & $\bullet$ & $\bullet$ & $\bullet$& $\bullet$& $\bullet$& $\bullet$& $\bullet$& $\bullet$ & \\ 
$N$ M5 & $\bullet$ & $\bullet$ & $\bullet$ & $\bullet$ & $\bullet$ & $\bullet$ \\
$\mathbb C^2 /\mathbb D_k$ & $\bullet$ & $\bullet$ & $\bullet$ & $\bullet$ & $\bullet$ & $\bullet$ & $\circ$ & $\circ$ & $\circ$ & $\circ$ &  $\bullet$ \\
\end{tabular}
\ee
Dualizing to Type I$'$ one obtains the following IIA brane system
\be\label{eq:IIAbrain-b}
\begin{tabular}{c|ccccccccccc}
&0 & 1 & 2 & 3 & 4 & 5 & 6 & 7 & 8  & 10\\
\hline
O8-D8 & $\bullet$ & $\bullet$ & $\bullet$ & $\bullet$ & $\bullet$& $\bullet$& $\bullet$& $\bullet$& $\bullet$&  & \\ 
$M$ NS5 + ON$^0$& $\bullet$ & $\bullet$ & $\bullet$ & $\bullet$ & $\bullet$ & $\bullet$ \\
$2k$ D6 + O6$^{\pm}$ & $\bullet$ & $\bullet$ & $\bullet$ & $\bullet$ & $\bullet$ & $\bullet$ & & & &  $\bullet$ \\
\end{tabular}
\ee
where with respect to the previous section the main difference is the presence of an $O6^\pm$ plane parallel to the stack of D6s. The latter changes sign across NS5 branes, which is compatible with the fractionalization of M5 branes. Moreover, since the $O6^\pm$ end up intersecting the O8$^-$ planes at a point, these systems also have ON$^0$ branes parallel to the NS5 brane stacks, localized at the point of intersection.  While some of the rules to manipulate these diagrams are well known (see e.g. \cite{Hanany:1997gh,Brunner:1997gf,Hanany:1999sj}), in some cases there are some interesting predictions from F-theory. For instance, from the results in \cite{Frey:2018vpw} we can read off the theory for a fractionalized M9-M5 system along a $D_4$ singularity for a choice of $\boldsymbol{\mu}:\mathbb D_4 \to E_8$ such that $F(\boldsymbol{\mu}) = E_8$. It is
\be
\mathcal T(\boldsymbol{\mu},\mathfrak g = \mathfrak{so}_8) = [\fe_8] \, \, 1 \,\, 2 \,\, \overset{\mathfrak{su}_2}{2}\,\, \overset{\mathfrak{g_2}}{3} \,\, 1 \,\, \overset{\mathfrak{so}_8}{4} \,\, 1\,\,[\mathfrak{so}_8]\,.
 \ee
We clearly see that the above tensor branch is slightly non-perturbative in nature. After fission we obtain
\be
\mathcal K_N(E_8,E_8;\mathfrak{so}_8) = [\fe_8] \, 1 \, 2 \, \overset{\mathfrak{su}_2}{2}\,\, \overset{\mathfrak{g_2}}{3} \,\, 1 \,\, \overset{\mathfrak{so}_8}{4} \,\, 1 \,\, \textcolor{red}{\overset{\mathfrak{so}_8}{4} } \, \underbrace{1 \, \overset{\mathfrak{so}_8}{4}\, 1\, \overset{\mathfrak{so}_8}{4} \, \cdots\, \overset{\mathfrak{so}_8}{4}\,1}_{\text{gauged }\mathcal T_{N-2}(\mathfrak{so}_8,\mathfrak{so}_8)}\, \textcolor{red}{\overset{\mathfrak{so}_8}{4}} \, 1\, \overset{\mathfrak{so}_8}{4} \, 1 \, \overset{\mathfrak{g_2}}{3}\, \overset{\mathfrak{su}_2}{2} \, 2 \,1 \, [\fe_8]
\ee
where a total of $N+1$ gauge groups of type $\mathfrak{so}_8$ are featured. 
The LS charge of this model is
\be
(\underbrace{1,1,1,1,2,1,2}_{\mathcal{T}(\fe_8,\mathfrak{so}_8)},\textcolor{red}{1},\underbrace{2,1,\dots,1,2}_{\mathcal T_{N-2}(\mathfrak{so}_8,\mathfrak{so}_8)},\textcolor{red}{1},\underbrace{2,1,2,1,1,1,1}_{\mathcal{T}(\fe_8,\mathfrak{so}_8)})
\ee
And therefore the R-symmetry 2-group structure constant for this theory is
\be
\begin{aligned}
\widehat{\kappa}_R &= 18 + 18 + 6 + 6 + 6(N-3) + 2(N-2) \\ &= 8N + 26
\end{aligned}
\ee
which correctly matches the one computed on the T-dual side. A more systematic analysis of the D-type cases can be achieved with similar methods. We will report on the results in future work on the topic.

\section{Heterotic instantons on exceptional singularities}\label{exceptional}

\begin{table}
\begin{center}
\begin{tabular}{c||c|c}
$\mathfrak{g}$ & $\mathcal T_1(\mathfrak{g},\mathfrak{g})$ & $\xymatrix{\ar@{-}[r]^{\mathfrak{g}}&}$\\
&\\
\hline
\hline
&\\
$E_6$ & $[\fe_6]\, 1 \, {\overset{\mathfrak{su}_3}{3}}\, 1  \, [\fe_6] $ & $\overset{\mathfrak{e_6}}{6}$\\
&\\
\hline
&\\
$E_7$ & $ [\fe_7]\, 1 \, {\overset{\mathfrak{su}_2}{2}}\,  {\overset{\mathfrak{so}_7}{3}}\,  {\overset{\mathfrak{su}_2}{2}}\, 1  \, [\fe_7] $& $\overset{\mathfrak{e_7}}{8}$\\
&\\
\hline
&\\
$E_8$ & $ [\fe_8]\, 1 \,  2 \, {\overset{\mathfrak{su}_2}{2}}\, {\overset{\mathfrak{g}_2}{3}}\, 1 \, {\overset{\mathfrak{f}_4}{5}}\, 1 \,  {\overset{\mathfrak{g}_2}{3}}\,  {\overset{\mathfrak{su}_2}{2}}\, 2 \,  1 \, [\fe_8]$ & $\overset{\mathfrak{e_8}}{12}$\\
&\\
\hline
\end{tabular}
\end{center}
\caption{The rank one conformal matter theories.}\label{tab:confmat}
\end{table}

The theories governing Heterotic $Spin(32)/\mathbb Z_2$ instantonic LSTs can be determined again by orbifolding techinques. For exceptional singularities, however, there are no (known) dual Type I$'$ realizations. The resulting theories have been determined by Intriligator and Blum for all the models with $\widetilde{\mathsf{w}}_2 = 0$. In this section we will therefore focus on the $E_8\times E_8$ side of the duality.

\medskip

The theories governing fractional heterotic instantons on exceptional ALE singularities can be described exploiting the results on the 6d SCFTs of type $\mathcal T(\boldsymbol{\mu},E_8)$  and the conformal matter theories $\mathcal T_N(E_s,E_s)$  that can be found in references \cite{DelZotto:2014hpa,Heckman:2015bfa,Frey:2018vpw}.

\begin{table}
\begin{center}
\renewcommand{\arraystretch}{0.8}
\begin{tabular}{c|c|l}
$\mathfrak{g} = \mathfrak{e}_s$ & $F(\boldsymbol{\mu})$ & $\mathcal T(F(\boldsymbol{\mu}),\mathfrak{g})$\\
&&\\
\hline
\hline
&&\\
$ \mathfrak{e}_6$ & $E_6$ & $[\fe_6] \, 1 \, {\overset{\mathfrak{su}_3}{3}}\, 1 \, {\overset{\mathfrak{f}_4}{5}}\, 1 \,  {\overset{\mathfrak{su}_3}{3}} \,1 \,\overset{\mathfrak{e}_6}{6}\,1 \,  {\overset{\mathfrak{su}_3}{3}} \,1 \,  [\fe_6]$ \\
&&\\
 & $E_6 \times U(1)$ & $[\fe_6] \, 1 \, {\overset{\mathfrak{su}_3}{3}}\, 1 \, \underset{[N_f =1]}{\overset{\mathfrak{e}_6}{5}}\, 1 \,  {\overset{\mathfrak{su}_3}{3}} \,1 \, \,\overset{\mathfrak{e}_6}{6}\,1 \,  {\overset{\mathfrak{su}_3}{3}} \,1 \,[\fe_6]$\\
&&\\
& $E_6 \times SU(3)$ & $[\fe_6] \, 1 \, {\overset{\mathfrak{su}_3}{3}}\, 1 \, \underset{1 \atop[\fsu_3]}{\overset{\mathfrak{e}_6}{6}}\, 1 \,  {\overset{\mathfrak{su}_3}{3}} \,1 \, [\fe_6]$\\
&&\\
& $E_7$ & $ [\fe_7]\, 1 \, {\overset{\mathfrak{su}_2}{2}}\, {\overset{\mathfrak{g}_2}{3}}\, 1 \, {\overset{\mathfrak{f}_4}{5}}\, 1 \, {\overset{\mathfrak{su}_3}{3}} \,1 \, \,\overset{\mathfrak{e}_6}{6}\,1 \,  {\overset{\mathfrak{su}_3}{3}} \,1 \,[\fe_6] $ \\
&&\\
& $E_8$ & $ [\fe_8]\, 1 \, 2\,  {\overset{\mathfrak{su}_2}{2}}\, {\overset{\mathfrak{g}_2}{3}}\, 1 \, {\overset{\mathfrak{f}_4}{5}}\, 1 \, {\overset{\mathfrak{su}_3}{3}} \,1 \,\,\overset{\mathfrak{e}_6}{6}\,1 \,  {\overset{\mathfrak{su}_3}{3}} \,1 \, [\fe_6] $\\
&&\\
\hline
&&\\
$ \mathfrak{e}_7$ & $E_6$ & $[\fe_6] \, 1 \, {\overset{\mathfrak{su}_3}{3}}\, 1 \, {\overset{\mathfrak{f}_4}{5}}\, 1 \, {\overset{\mathfrak{g}_2}{3}} \, {\overset{\mathfrak{su}_2}{2}}\, 1 \,\,\overset{\mathfrak{e}_7}{7}\,   \, 1 \, {\overset{\mathfrak{su}_2}{2}}\,  {\overset{\mathfrak{so}_7}{3}}\,  {\overset{\mathfrak{su}_2}{2}}\, 1\, [\fe_7]  $ \\
&&\\
& $E_6'$ & $[\fe_6] \, 1 \, {\overset{\mathfrak{su}_3}{3}}\, 1 \, {\overset{\mathfrak{e}_6}{6}}\, 1 \,  {\overset{\mathfrak{su}_2}{2}} \, {\overset{\mathfrak{so}_7}{3}} \, {\overset{\mathfrak{su}_2}{2}}\, 1 \,\,\overset{\mathfrak{e}_7}{7}\,   \, 1 \, {\overset{\mathfrak{su}_2}{2}}\,  {\overset{\mathfrak{so}_7}{3}}\,  {\overset{\mathfrak{su}_2}{2}}\, 1\, [\fe_7]  $  \\
&&\\
& $E_7$ & $[\fe_7] \, 1 \, {\overset{\mathfrak{su}_2}{2}}\,  {\overset{\mathfrak{g}_2}{3}} \, 1 \, {\overset{\mathfrak{f}_4}{5}}\, 1 \, {\overset{\mathfrak{g}_2}{3}} \, {\overset{\mathfrak{su}_2}{2}}\, 1 \,\overset{\mathfrak{e}_7}{7}\,   \, 1 \, {\overset{\mathfrak{su}_2}{2}}\,  {\overset{\mathfrak{so}_7}{3}}\,  {\overset{\mathfrak{su}_2}{2}}\, 1\, [\fe_7] $ \\
&&\\
& $E_7'$ & $[\fe_7] \, 1 \, {\overset{\mathfrak{su}_2}{2}}\,  {\overset{\mathfrak{so}_7}{3}}\,  {\overset{\mathfrak{su}_2}{2}}\, 1 \, \underset{[N_f=1/2]}{\overset{\mathfrak{e}_7}{7}}\,   \, 1 \, {\overset{\mathfrak{su}_2}{2}}\,  {\overset{\mathfrak{so}_7}{3}}\,  {\overset{\mathfrak{su}_2}{2}}\, 1\, [\fe_7]$\\
&&\\
& $E_7 \times SU(2)$ & $[\fe_7] \, 1 \, {\overset{\mathfrak{su}_2}{2}}\,  {\overset{\mathfrak{so}_7}{3}}\,  {\overset{\mathfrak{su}_2}{2}}\, 1 \, \underset{1 \atop [\fsu_2]}{\overset{\mathfrak{e}_7}{8}}\,   \, 1 \, {\overset{\mathfrak{su}_2}{2}}\,  {\overset{\mathfrak{so}_7}{3}}\,  {\overset{\mathfrak{su}_2}{2}}\, 1\, [\fe_7]$\\
&&\\
& $E_8$ & $ [\fe_8]\, 1 \, 2\,  {\overset{\mathfrak{su}_2}{2}}\, {\overset{\mathfrak{g}_2}{3}}\, 1 \, {\overset{\mathfrak{f}_4}{5}}\, 1 \,  {\overset{\mathfrak{g}_2}{3}} \, {\overset{\mathfrak{su}_2}{2}}\, 1 \, \,\overset{\mathfrak{e}_7}{7}\,   \, 1 \, {\overset{\mathfrak{su}_2}{2}}\,  {\overset{\mathfrak{so}_7}{3}}\,  {\overset{\mathfrak{su}_2}{2}}\, 1\,[\fe_7]$\\
&&\\
\hline
&&\\
$ \mathfrak{e}_8$ & $E_6$ & $[\fe_6] \, 1 \, {\overset{\mathfrak{su}_3}{3}}\, 1 \, {\overset{\mathfrak{f}_4}{5}}\, 1 \, {\overset{\mathfrak{g}_2}{3}} \, {\overset{\mathfrak{su}_2}{2}}\, 2 \,1 \, \overset{\mathfrak{e}_8}{\underset{1}{(12)}} \, 1 \,  2 \, {\overset{\mathfrak{su}_2}{2}}\, {\overset{\mathfrak{g}_2}{3}}\, 1 \, {\overset{\mathfrak{f}_4}{5}}\, 1 \,  {\overset{\mathfrak{g}_2}{3}}\,  {\overset{\mathfrak{su}_2}{2}}\, 2 \,  1 \,  [\fe_8]$ \\
&&\\
 & $E_6'$ & $[\fe_6] \, 1 \, {\overset{\mathfrak{su}_3}{3}}\, 1 \, \overset{\mathfrak{e}_6}{6} \, 1 \, {\overset{\mathfrak{su}_3}{3}}\, 1 \, {\overset{\mathfrak{f}_4}{5}}\, 1 \, {\overset{\mathfrak{g}_2}{3}} \, {\overset{\mathfrak{su}_2}{2}}\, 2 \,1 \,\, [\fe_8]$ \\
&&\\
& $E_7$ & $[\fe_7] \, 1 \, {\overset{\mathfrak{su}_2}{2}}\,  {\overset{\mathfrak{g}_2}{3}} \, 1 \, {\overset{\mathfrak{f}_4}{5}}\, 1 \, {\overset{\mathfrak{g}_2}{3}} \, {\overset{\mathfrak{su}_2}{2}}\, 2 \,1 \, \overset{\mathfrak{e}_8}{\underset{1}{(12)}} \, 1 \,  2 \, {\overset{\mathfrak{su}_2}{2}}\, {\overset{\mathfrak{g}_2}{3}}\, 1 \, {\overset{\mathfrak{f}_4}{5}}\, 1 \,  {\overset{\mathfrak{g}_2}{3}}\,  {\overset{\mathfrak{su}_2}{2}}\, 2 \,  1 \,  [\fe_8]$ \\
&&\\
& $E_7'$ & $[\fe_7] \, 1 \,  {\overset{\mathfrak{su}_2}{2}}\,  {\overset{\mathfrak{so}_7}{3}}\,  {\overset{\mathfrak{su}_2}{2}}\, 1 \, \overset{\mathfrak{e}_7}{8} \, 1 \, {\overset{\mathfrak{su}_2}{2}}\, {\overset{\mathfrak{g}_2}{3}}\, 1 \, {\overset{\mathfrak{f}_4}{5}}\, 1 \,  {\overset{\mathfrak{g}_2}{3}}\,  {\overset{\mathfrak{su}_2}{2}}\, 2 \,  1 \,\, [\fe_8]$ \\
&&\\
& $E_8$ & $[\fe_8] \, 1 \, 2\,  {\overset{\mathfrak{su}_2}{2}}\,  {\overset{\mathfrak{g}_2}{3}} \, 1 \, {\overset{\mathfrak{f}_4}{5}}\, 1 \, {\overset{\mathfrak{g}_2}{3}} \, {\overset{\mathfrak{su}_2}{2}}\, 2 \,1 \, \overset{\mathfrak{e}_8}{\underset{1}{(12)}} \, 1 \,  2 \, {\overset{\mathfrak{su}_2}{2}}\, {\overset{\mathfrak{g}_2}{3}}\, 1 \, {\overset{\mathfrak{f}_4}{5}}\, 1 \,  {\overset{\mathfrak{g}_2}{3}}\,  {\overset{\mathfrak{su}_2}{2}}\, 2 \,  1 \,  [\fe_8]$  \\
&&\\
\hline
\end{tabular}
\end{center}
\caption{Choices of $\boldsymbol{\mu}$ leading to exceptional symmetries along $\mathfrak{e}_{6,7,8}$-type ALE spaces.}\label{tab:flatty}
\end{table}

The theories $\mathcal T_N(E_s,E_s)$ can be understood as fusions of the corresponding rank one theories, by the recursive formula
\be
\mathcal T_N(E_s,E_s) = \mathcal T_{N-1}(E_s,E_s) \xymatrix{\ar@{-}[r]^{E_s}&} \mathcal T_1(E_s,E_s)
\ee
In table $\ref{tab:confmat}$ we summarise the data of exceptional conformal matter theories we will need in this section. As an example
\be
\begin{aligned}
\mathcal T_2(E_6,E_6) &= \mathcal T_1(E_6,E_6) \xymatrix{\ar@{-}[r]^{E_6}&}  T_1(E_6,E_6) \\
&=  [\fe_6]\, 1 \, {\overset{\mathfrak{su}_3}{3}}\, 1 \, \textcolor{red}{{\overset{\mathfrak{e}_6}{6}}} \, 1 \, {\overset{\mathfrak{su}_3}{3}}\, 1 \, [\fe_6]\,.
\end{aligned}
\ee
The models $\mathcal T(\boldsymbol{\mu},E_8)$ have been studied in details in reference \cite{Frey:2018vpw}. In this paper we focus on the choices of $\boldsymbol{\mu}_a$ for which $F(\boldsymbol{\mu}_a)$ has at least one exceptional factor. We summarise the corresponding theories in Table \ref{tab:flatty}. Given our definitions above, the 6d (1,0) little string theory of $N$ heterotic $E_8 \times E_8$ instantons probing an $E$-type singularity is then given by
\be
\xymatrix{\mathcal T(\boldsymbol{\mu}_1,\mathfrak{e_s})\ar@{-}[r]^{\mathfrak{e_s}}&\mathcal T_{N-2}(\mathfrak{e_s},\mathfrak{e_s})\ar@{-}[r]^{\mathfrak{e_s}}&\mathcal T(\boldsymbol{\mu}_2,\mathfrak{e_s})}
\ee
Exploiting these results and the fusion operations of equation \eqref{eq:avheterotto} it is straightforward to determine the LSTs of interest, which we will denote 
\be
\mathcal K_N(F(\boldsymbol{\mu}_1),F(\boldsymbol{\mu}_2);\mathfrak g = \mathfrak e_s)\,.
\ee

\medskip

As an example, consider the case of an $E_6$ ALE singularity and take the heterotic instantonic LSTs with flavor symmetries $F_1 = E_6 \times U(1)$ and $F_2 = E_7$. We obtain that $\mathcal K_N(E_6 \times U(1),E_7;\mathfrak g = \mathfrak e_6)$ has generalized quiver
\be
[\fe_6] \, 1 \, {\overset{\mathfrak{su}_3}{3}}\, 1 \, \underset{[N_f=1]}{\overset{\mathfrak{e}_6}{5}}\, 1 \,  {\overset{\mathfrak{su}_3}{3}} \,1 \, \textcolor{red}{\overset{\mathfrak{e}_6}{6}} \, \underbrace{1 \, 3 \, 1 \, \overset{\mathfrak{e}_6}{6} \,1 \, 3 \, 1 \cdots\, 1 \, 3 \, 1  \,\overset{\mathfrak{e}_6}{6} \, 1 \, {\overset{\mathfrak{su}_3}{3}} \,1}_{\mathcal{T}_{N-2}(\mathfrak{e}_6,\mathfrak{e}_6)} \,  \textcolor{red}{\overset{\mathfrak{e}_6}{6}} \,1 \, {\overset{\mathfrak{su}_3}{3}} \,1\,\overset{\mathfrak{e}_6}{6}  \, 1 \, {\overset{\mathfrak{su}_3}{3}} \,1  \, {\overset{\mathfrak{f}_4}{5}}\,  1  \, {\overset{\mathfrak{g}_2}{3}}\, {\overset{\mathfrak{su}_2}{2}}\, 1 \, [\fe_7]\,.
\ee
where again we have indicated in red the fission nodes.

\medskip

It is straightforward to extend our results to a more systematic explorations of all possible other cases corresponding to more general choices of $\boldsymbol{\mu}_a: \Gamma_{\mathfrak{g}} \to E_8$, obtaining by patching together two copies of the models discussed in \cite{Frey:2018vpw}. We leave a systematic study of these examples for the future.

\medskip

As a further check for this characterization of this class of models, we now turn to the corresponding T-dualities. We stress that for the cases of exceptional singularities, there are no dual brane configurations, and to prove these T-dualities one needs to exploit F-theory. In the section below we show that the corresponding 2-group structure constants and Coulomb branch dimensions for the theories we propose do indeed match with the ones of known $Spin(32)/\mathbb Z_2$ instantons.

\subsection{The case of $\mathfrak g = \mathfrak{e}_6$}
From the analysis by Intriligator and Blum \cite{Blum:1997mm} it follows that the $Spin(32)/\mathbb Z_2$ Heterotic instantons which are compatible with a vector structure all have generalized quivers of the form\footnote{\ Our conventions for the integers $v_i$ and $w_i$ differs slightly from the ones in \cite{Blum:1997mm}.}
\be
\begin{matrix}
\underset{[\mathfrak{so}_{2w_1}]}{\overset{\mathfrak{sp}_{v_1}}{1}} \,\, \underset{[\mathfrak{sp}_{w_2}]}{\overset{\mathfrak{so}_{2v_2}}{4}} \,\, \underset{[\mathfrak{so}_{2w_3}]}{\overset{\mathfrak{sp}_{v_3}}{1}} \,\, \underset{[\mathfrak{u}_{w_4}]}{\overset{\mathfrak{su}_{v_4}}{2}} \,\, \underset{[\mathfrak{u}_{w_4}]}{\overset{\mathfrak{su}_{v_5}}{2}}
\end{matrix}
\ee
which have LS charge
\be
\vec{\ell}_{LS}=(1,1,3,2,1)
\ee
and therefore
\be
\widehat{\kappa}_R = 2 + v_1 + 3 v_3 + 2 v_2 + 2 v_4 + v_5
\ee
while the corresponding Coulomb branch dimension is
\be
d_{CB} = 2 + v_1 + v_2 + v_3 + v_4 + v_ 5 
\ee
The coefficients $v_i$ above are a function of the $w_i$ which are in turn determined by the choice of $\boldsymbol{\lambda}$. We refer our readers to \cite{Blum:1997mm} for the details of the dictionary, which we use here only to provide the examples that we need to give evidence for the structure of the Heterotic $E_8\times E_8$ instantons we are constructing. We report in Table \ref{tab:E6cases} our results for the models of the form
\be
\mathcal K_N(E_r,E_s;\mathfrak{e}_6)= \xymatrix{\mathcal T(E_r,\mathfrak{e_6})\ar@{-}[r]^{\mathfrak{e_6}}&\mathcal T_{N-2}(\mathfrak{e_6},\mathfrak{e_6})\ar@{-}[r]^{\mathfrak{e_6}}&\mathcal T(E_s,\mathfrak{e_6})}
\qquad s,r \in \{6,7,8\}
\ee
where the theories $\mathcal T(E_r,\mathfrak{e_6})$ can be read off from Table \ref{tab:flatty}. For each of the proposed theories we find at least one model among the Heterotic $Spin(32)/\mathbb Z_2$ instantonic LSTs of type $\widetilde{K}_N(\boldsymbol{\lambda};\mathfrak{g}=\mathfrak{e}_6)$ which satisfies the necessary conditions to be a T-dual theory.

\begin{table}
\begin{center}
\begin{tabular}{cc||cc||l}
$F(\boldsymbol{\mu}_1)$ & $F(\boldsymbol{\mu}_2)$ & $\widehat{\kappa}_R$ & $d_{CB}$ & $\widetilde{K}_N(\boldsymbol{\lambda};\mathfrak{g}=\mathfrak{e}_6)$ dual\phantom{$\Big|$}\\
\hline
&&&&\\
$E_8$ & $E_8$ & $74 + 24N$ & $42+12N$ & \SEsix{[\mathfrak{so}_{32}]}{}{}{}{}{{10 + N}{24 + 4 N}{6 + 3 N}{8 + 4 N}{4 + 2 N}}\\
&&&&\\
$E_8$ & $E_7$ & $73 + 24 N$ & $41 + 12 N$ & \SEsix{[\mathfrak{so}_{28}]}{[\mathfrak{sp}_1]}{}{}{}{{9 + N}{24 + 4 N}{6 + 3 N}{8 + 4 N}{4 + 2 N}}\\
&&&&\\
$E_7$ & $E_7$ & $72 + 24 N$ & $40 + 12N$ & \SEsix{[\mathfrak{so}_{24}]}{[\mathfrak{sp}_2]}{}{}{}{{8 + N}{24 + 4 N}{6 + 3 N}{8 + 4 N}{4 + 2 N}}\\
&&&&\\
$E_8$ & $E_6$ & $70 + 24 N$ & $39 + 12 N$ & \SEsix{[\mathfrak{so}_{26}]}{}{[\mathfrak{so}_2]}{}{}{{8 + N}{22 + 4 N}{6 + 3 N}{8 + 4 N}{4 + 2 N}}\\
&&&&\\
$E_7$ & $E_6$ & $69 + 24 N$ & $38 + 12 N$ & \SEsix{[\mathfrak{so}_{22}]}{[\mathfrak{sp_1}]}{[\mathfrak{so}_2]}{}{}{{7 + N}{22 + 4 N}{6 + 3 N}{8 + 4 N}{4 + 2 N}}\\
&&&&\\
$E_6$ & $E_6$ & $66 + 24 N$ & $36 + 12 N$ & \SEsix{[\mathfrak{so}_{20}]}{}{[\mathfrak{so}_{4}]}{}{}{{6 + N}{20 + 4 N}{6 + 3 N}{8 + 4 N}{4 + 2 N}}\\
&&&&\\
\end{tabular}
\caption{T-dual theories for the $\mathcal K_N(F(\boldsymbol{\mu}_1),F(\boldsymbol{\mu}_2),\mathfrak{e}_6)$ LSTs. Notice that the Coulomb branch dimensions are increasing by units of $h^\vee_{E_6} = 12$.}\label{tab:E6cases}
\end{center}
\end{table}

\subsection{The case of $\mathfrak g = \mathfrak{e}_7$}
In this section we consider the $E_8 \times E_8$ LSTs along the $\mathfrak{e}_7$ singularity. We focus on models of the form
\be
\mathcal{K}_N(E_r,E_s; \mathfrak{e}_7) = \xymatrix{\mathcal T(E_r,\mathfrak{e_7})\ar@{-}[r]^{\mathfrak{e_7}}&\mathcal T_{N-2}(\mathfrak{e_7},\mathfrak{e_7})\ar@{-}[r]^{\mathfrak{e_7}}&\mathcal T(E_s,\mathfrak{e_7})}\qquad s,r \in \{6,7,8\}
\ee
Sometimes different embeddings
\be
\boldsymbol{\mu},\boldsymbol{\mu}':\Gamma_{\mathfrak{e}_7} \to E_8
\ee
are such that
\be
F(\boldsymbol{\mu}) = F(\boldsymbol{\mu}') = E_s
\ee
If that is the case, to distinguish the corresponding $M9-M5$ theories, we denote them $\mathcal T(E_s,\mathfrak{e}_7)$, and $\mathcal T(E_s',\mathfrak{e}_7)$ in Table \ref{tab:flatty}. 

\medskip

The generalized 6d quivers for the $\widetilde{\mathcal{K}}_N(\boldsymbol{\lambda};\mathfrak{g} = \mathfrak e_7)$ have the form
\be
\begin{matrix}
&&&[\mathfrak{so_{2w_8}}]\atop \qquad\\
&&&\overset{\mathfrak{sp}_{v_8}}{1}\\
[\mathfrak{so}_{2w_1}] \,\,\overset{\mathfrak{sp}_{v_1}}{1}&\underset{[\mathfrak{sp}_{w_2}]}{\overset{\mathfrak{so}_{2v_2}}{4}}&\underset{[\mathfrak{so}_{2w_3}]}{\overset{\mathfrak{sp}_{v_3}}{1}}&\underset{[\mathfrak{sp}_{w_4}]}{\overset{\mathfrak{so}_{2v_4}}{4}}&\underset{[\mathfrak{so}_{2w_5}]}{\overset{\mathfrak{sp}_{v_5}}{1}}&\underset{[\mathfrak{sp}_{w_6}]}{\overset{\mathfrak{so}_{2v_6}}{4}}&\overset{\mathfrak{sp}_{v_7}}{1} \,\,[\mathfrak{so}_{2w_7}] \\
\end{matrix}
\ee
where the $v_i$ are a function of the $w_i$, which in turn are encoded by $\boldsymbol{\lambda}$ as explained in details in \cite{Blum:1997mm}. From the generalized quiver above we read off the LS charge
\be 
\begin{matrix}
&&&2,\\
\vec{\ell}_{LS}=(1,&1,& 3,& 2,& 3,& 1,& 1)
\end{matrix}
\ee
and hence 
\be
\begin{aligned}
\widehat{\kappa}_R &=2+ v_1 + 2 v_2 + 3 v_3 + 4 v_4 + 3 v_5 + 2 v_6 + v_7 + 2 v_8\\
d_{CB}& = 7+ \sum_{i=1}^8 v_i
\end{aligned}
\ee
We find T-dual models with the same features for all the theories we propose. We have explicitly checked all possible combinations with exceptional symmetries, but the check is not that instructive. We  report some of our results in Table \ref{tab:E7cases}. It would be interesting to carry out a more systematic scan of these possibilities, along the lines of the analysis we have done for the $\mathbb C^2/\mathbb Z_2$ singularity. We expect that from one such systematic study several novel families of T-dualities will emerge.

\begin{table}
\hspace{-1.6cm}\begin{tabular}{cc||cc|c}
$F(\boldsymbol{\mu}_1)$ & $F(\boldsymbol{\mu}_2)$ & $\widehat{\kappa}_R$ & $d_{CB}$ & $\widetilde{K}_N(\boldsymbol{\lambda};\mathfrak{g}=\mathfrak{e}_7)$ dual\phantom{$\Big|$}\\
\hline
&&&&\\
$E_8$ &$E_8$ & $98 + 48N$ & $47 + 18 N$ & \SEse{[\mathfrak{so}_{32}]}{}{}{}{}{}{}{}{{24 + 4 N} {6 + 3 N}{16 + 8 N}{2 + 3 N}{8 + 4 N}{-2 + 
  N}{2 N}{10 + N}} \\
$E_7$ &$E_8$ & $97 + 48N$ & $46 + 18 N$ &\SEse{[\mathfrak{so}_{28}]}{[\mathfrak{sp}_1]}{}{}{}{}{}{}{{24 + 4 N}{6 + 3 N}{16 + 8 N}{2 + 3 N}{8 + 4 N}{-2 + N}{2 N}{9 + N}}\\
$E_7$ &$E_7$ & $96 + 48N$ & $45 + 18 N$ &\SEse{[\mathfrak{so}_{24}]}{[\mathfrak{sp}_2]}{}{}{}{}{}{}{{24 + 4 N}{6 + 3 N}{16 + 8 N}{2 + 3 N}{8 + 4 N}{-2 + N}{2 N}{8 + N}}\\
$E_6$ &$E_8$ & $94 + 48N$ & $44 + 18 N$ &\SEse{[\mathfrak{so}_{26}]}{}{[\mathfrak{so}_{2}]}{}{}{}{}{}{{22 + 4 N}{6 + 3 N}{16 + 8 N}{2 + 3 N}{8 + 4 N}{-2 + N}{2 N}{8 + N}}\\
$E_6$ &$E_7$ & $93 + 48N$ & $43 + 18 N$ &\SEse{[\mathfrak{so}_{22 }]}{ [\mathfrak{sp}_{1}]}{[\mathfrak{so}_{2}]}{}{}{}{}{}{{22 + 4 N}{6 + 3 N}{16 + 8 N}{2 + 3 N}{8 + 4 N}{-2 + N}{2 N}{7 + N}}\\
$E_6$ &$E_6$ & $90 + 48N$ & $41 + 18 N$ &\SEse{[\mathfrak{so}_{20}]}{}{[\mathfrak{so}_{4}]}{}{}{}{}{}{{20 + 4 N}{6 + 3 N}{16 + 8 N}{2 + 3 N}{8 + 4 N}{-2 + N}{2 N}{6 + N}}\\
$E_7'$ &$E_8$ &$73 + 48N$ & $37 + 18 N$ &\SEse{}{[\mathfrak{sp}_1]}{}{}{}{}{[\mathfrak{so}_{28}]}{}{{8 + 4 N}{1 + 3 N}{12 + 8 N}{4 + 3 N}{20 + 4 N}{
 8 + N}{-1 + 2 N}{-2 + N}}\\
 $E_7'$ &$E_7'$ &  $48 + 48N$ & $27 + 18 N$ &\SEse{}{[\mathfrak{sp}_2]}{}{}{}{}{[\mathfrak{so}_{24}]}{}{{8 + 4 N}{3 N}{8 + 8 N}{2 + 3 N}{16 + 4 N}{6 + N}{-2 + 2 N}{-2 + N}}\\
\end{tabular}
\caption{T-dual theories for selected $\mathcal K_N(F(\boldsymbol{\mu}_1),F(\boldsymbol{\mu}_2),\mathfrak{e}_7)$ LSTs. Notice that the Coulomb branch dimensions are increasing by units of $h^\vee_{E_7} = 18$.}\label{tab:E7cases}
\end{table}

\subsection{The case of $\mathfrak g = \mathfrak{e}_8$}

\begin{table}
\hspace{-1.5cm}\scalebox{0.9}{\hspace{-1cm} \begin{tabular}{c|c||c|c||c}
\phantom{\Big|}$F(\boldsymbol{\mu}_1)$ & $F(\boldsymbol{\mu}_2)$ & $\widehat{\kappa}_R$ & $d_{CB}$ & Dual  $\widetilde{\mathcal K}_N(\boldsymbol{\lambda};\mathfrak g= \mathfrak{e}_8)$ theory\\
\hline
\hline
&&&&\\
$E_8$ & $E_8$ & $122+120 N$ & $ 52 +30 N$ & \SEIT{}{}{}{}{}{}{}{{}{[\mathfrak{so}_{32}]}{{8 + 4 N}{4 N}{8 + 12 N}{2 + 5 N}{16 + 8 N}{6 + 3 N}{24 + 4 N}{-2 + 3 N}{10 + N}}}\\

$E_7$ & $E_8$ & $121+120 N$ & $ 51 +30 N$ & \SEIT{}{}{}{}{}{}{[\mathfrak{sp}_1]}{{}{[\mathfrak{so}_{28}]}{{8 + 4 N}{4 N}{8 + 12 N}{2 + 5 N}{16 + 8 N}{6 + 3 N}{24 + 4 N}{-2 + 3 N}{9 + N}}}\\

$E_7$ & $E_7$ & $120 + 120 N$ & $ 50 +30 N$ & \SEIT{}{}{}{}{}{}{[\mathfrak{sp}_2]}{{}{[\mathfrak{so}_{24}]}{{8 + 4 N}{4 N}{8 + 12 N}{2 + 5 N}{16 + 8 N}{6+ 3 N}{24 + 4N}{-2 + 3 N}{8 + N}}}\\

$E_6$ & $E_8$ & $118 +120 N$ & $ 49 +30 N$ & \SEIT{}{}{}{}{}{[\mathfrak{so}_2]}{}{{}{[\mathfrak{so}_{26}]}{{8 + 4 N}{4 N}{8 + 12 N}{2 + 5 N}{16 + 8 N}{6 + 3 N}{22 + 4 N}{-2 + 3 N}{8 + N}}}\\
  
$E_6$ & $E_7$ & $117 + 120N$ & $48 + 30N$ &\SEIT{}{}{}{}{}{[\mathfrak{so}_{2}]}{[\mathfrak{sp}_1]}{{}{[\mathfrak{so}_{22}]}{{8 + 4 N}{4 N}{8 + 12 N}{2 + 5 N}{16 + 8 N}{6 + 3 N}{22 + 4 N}{-2 + 3 N}{7 + N}}}\\

$E_6$ & $E_6$ & $114 + 120N$ & $46 + 30N$ &\SEIT{}{}{}{}{}{[\mathfrak{so}_4]}{}{{}{[\mathfrak{so}_{20}]}{{8 + 4 N}{4 N}{8 + 12 N}{2 + 5 N}{16 + 8 N}{6 + 3 N}{20 + 4 N}{-2 + 3 N}{6 + N}}}\\

$E_7'$ & $E_8$ & $73 +120 N$ & $ 38 +30 N$ & \SEIT{[\mathfrak{sp}_{1}]}{}{}{}{}{}{}{{}{[\mathfrak{so}_{28}]}{{8 + 4 N}{-1 + 4 N}{4 + 12 N}{5 N}{12 + 8 N}{4 +3 N}{20 + 4 N}{-3 + 3 N}{8 + N}}}\\

$E_7$ & $E_7'$ & $72 + 120 N$ & $ 37 +30 N$ & \SEIT{[\mathfrak{sp}_{1}]}{}{}{}{}{}{[\mathfrak{sp}_{1}]}{{}{[\mathfrak{so}_{24}]}{{8 + 4 N}{-1 + 4 N}{4 + 12 N}{5 N}{12 + 8 N}{4 + 3 N}{20 + 4 N}{-3 + 3 N}{7 + N}}}\\

$E_6'$ & $E_8$ & $46 +120 N$ & $ 30 +30 N$ & \SEIT{}{}{}{}{}{}{}{{[\mathfrak{so}_{2}]}{[\mathfrak{so}_{26}]}{{6 + 4 N}{-2 + 4 N}{2 + 12 N}{-1 + 5 N}{10 + 8 N}{3 + 3 N}{18 + 4 N}{-3 + 3 N}{7 + N}}}\\
\end{tabular}}
\caption{ $\mathfrak{g} = \mathfrak{e}_8$ cases. Notice that again $d_{CB}$ is increasing in steps of $h^\vee_{\mathfrak {e}_8} = 30$.}\label{tab:e8casesss}
\end{table}

\begin{table}
\hspace{-2cm}\scalebox{0.9}{\begin{tabular}{c|c||c|c||c}
\phantom{\Big|}$F(\boldsymbol{\mu}_1)$ & $F(\boldsymbol{\mu}_2)$ & $\widehat{\kappa}_R$ & $d_{CB}$ & Dual $\widetilde{\mathcal K}_N(\boldsymbol{\lambda};\mathfrak g= \mathfrak{e}_8)$ theory\\
\hline
\hline
&&&&\\

$E_6$ & $E_7'$ & $69 + 120N$ & $35 + 30N$ &\SEIT{[\mathfrak{sp}_{1}]}{}{}{}{}{[\mathfrak{so}_2]}{}{{}{[\mathfrak{so}_{22}]}{{8 + 4 N}{-1 + 4 N}{4 + 12 N}{5 N}{12 + 8 N}{4 + 3 N}{18 + 4 N}{-3 + 3 N}{6 + N}}}\\

$E_6'$ & $E_7$ & $45 + 120N$ & $29 + 30N$ &\SEIT{}{}{}{}{}{}{[\mathfrak{sp}_1]}{{[\mathfrak{so}_2]}{[\mathfrak{so}_{22}]}{{6 + 4 N}{-2 + 4 N}{2 + 12 N}{-1 + 5 N}{10 + 8 N}{3 + 3 N}{18 + 4 N}{-3 + 3 N}{6 + N}}}\\

$E_6'$ & $E_6$ & $42 + 120M$ & $27 + 30N$ &\SEIT{}{}{}{}{}{[\mathfrak{so}_2]}{}{{[\mathfrak{so}_2]}{[\mathfrak{so}_{20}]}{{6 + 4 N}{-2 + 4 N}{2 + 12 N}{-1 + 5 N}{10 + 8 N}{3 + 3 N}{16 + 4 N}{-3 + 3 N}{5 + N}}}\\

$E_7'$ & $E_7'$ & $24 + 120 N$ & $ 24 +30 N$ & \SEIT{[\mathfrak{sp}_2]}{}{}{}{}{}{}{{}{[\mathfrak{so}_{24}]}{{8 + 4 N}{-2 + 4 N}{12 N}{-2 + 5 N}{8 + 8 N}{2 + 3 N}{16 + 4 N}{-4 + 3 N}{6 + N}}}\\

$E_6'$ & $E_7'$ & $-3 + 120N$ & $16 + 30N$ &\SEIT{[\mathfrak{sp}_1]}{}{}{}{}{}{}{{[\mathfrak{so}_2]}{[\mathfrak{so}_{22}]}{{6 + 4 N}{-3 + 4 N}{-2 + 12 N}{-3 + 5 N}{6 + 8 N}{1 + 3 N}{14 + 4 N}{-4 + 3 N}{5 + N}}}\\

$E_6'$ & $E_6'$ & $-30 + 120N$ & $8 + 30N$ &\SEIT{}{}{}{}{}{}{}{{[\mathfrak{so}_{4}]}{[\mathfrak{so}_{20}]}{{4 + 4 N}{-4 + 4 N}{-4 + 12 N}{-4 + 5 N}{4 + 8 N}{3 N}{12 + 4 N}{-4 + 3 N}{4 + N}}}\\
\end{tabular}}
\caption{ $\mathfrak{g} = \mathfrak{e}_8$ cases -- continued.}
\end{table}

The analysis of the $E_8 \times E_8$ instantonic LSTs along the $\mathfrak{e}_8$ singularity proceeds in a similar way. For these models, the Intriligator Blum duals have the form
\be
\text{ \SEIT{[\mathfrak{sp}_{w_7}]}{[\mathfrak{so}_{2w_6}]}{[\mathfrak{sp}_{w_5}]}{[\mathfrak{so}_{2w_4}]}{[\mathfrak{sp}_{w_3}]}{[\mathfrak{so}_{2w_2}]}{[\mathfrak{sp}_{w_1}]}{{[\mathfrak{so}_{2w_8}]}{[\mathfrak{so}_{2w_9}]}{{4 + 4 M}{-4 + 4 M}{2v_3}{v_4}{2v_5}{v_6}{2v_7}{v_8}{v_9}}}}
\ee
where the $v_i$ are a function of the $w_i$, which in turn are encoded by $\boldsymbol{\lambda}$ as explained in details in \cite{Blum:1997mm}. The corresponding LS charge is
\be
\begin{matrix}
&& 3,\\
\vec{\ell}_{LS}=(1, & 4, & 3, & 5, & 2, & 3, & 1, & 1)
\end{matrix}
\ee
from which one can easily extract the value of $\widehat{\kappa}_R$ for this class of theories. Again exploiting the known T-duals we can confirm our results on the $E_8\times E_8$ side --- see Table \ref{tab:e8casesss} for some results.

\section{Proposal for an $a$-Theorem for 6d LSTs}\label{sec:atheorem}

We conclude this paper with an interesting remark about the structure of the LSTs we have discussed. All the orbi-instanton theories labeled by pairs $(\Gamma_\mathfrak{g},\boldsymbol{\mu})$ are connected via Higgs branch RG flows. This triggers Higgs branch RG flows involving 6d LSTs of the type we consider in this paper. More precisely, whenever there is an RG flow of the form
\be
\mathcal T(\boldsymbol{\mu}, \mathfrak g_\Gamma) \xrightarrow{\qquad RG \qquad}  \mathcal T(\widetilde{\boldsymbol{\mu}}, \mathfrak g_\Gamma)
\ee
we expect that it induces an RG flow among the corresponding LSTs
\be\label{eq:LSTRG}
\mathcal K_N(\boldsymbol{\mu},\boldsymbol{\mu}^\prime; \mathfrak{g}) \xrightarrow{\qquad RG \qquad}  \mathcal K_N(\widetilde{\boldsymbol{\mu}},\boldsymbol{\mu}^\prime; \mathfrak{g})\,.
\ee
RG flows among orbi-instanton theories have been widely studied, and therefore one can relatively simply chart such RG flows \cite{Heckman:2016ssk,Heckman:2018pqx,Fazzi:2022hal}.

\medskip

It is natural to ask whether in this context we can give evidence for the existence of an $a$-Theorem for 6d Little String Theories. This theorem must be non-standard because $a$-Theorems typically involve the coefficient of the Euler density of the Weyl anomalies for the trace of the stress energy tensor $T_{\mu\nu}$ 
\be
\langle T^\mu_\mu \rangle \sim a E_d + \cdots 
\ee
in presence of a background metric,\footnote{\ The $\cdots$ indicate the possible presence of other dimension-dependent Weyl invariants of degree $d$.} but LSTs do not have a well defined stress energy tensor because of T-duality. Hence, a question arises naturally: is there a function which is monotonically decreasing along LST RG flows? For 6d SCFTs and $a$-theorem can be argued for \cite{Elvang:2012st,Cordova:2015fha}, hence it is possible that there is an extension of such a theorem which holds for 6d LSTs.

\medskip

Computing $\widehat\kappa_R$ for theories related by RG flows of the form in equation \eqref{eq:LSTRG}, one can indeed see that
\be
\widehat\kappa_R(\textsc{lst}_{\textsc{uv}}) > \widehat\kappa_R(\textsc{lst}_{\textsc{ir}})
\ee
thus indicating that $\widehat\kappa_R$ is a quantity which is decreasing along RG flows.

\medskip 

All examples we have considered in this paper that are connected by RG flows indeed have this feature -- notice that the RG flows among LSTs are never of the tensor-branch kind, because removing a single tensor from a 6d LST would give rise to a 6d SCFT.

\section*{Acknowledgements}

We thank Iñaki García Etxebarria, Julius Grimminger, Amihay Hanany,  and Cumrun Vafa for discussions. The work of MDZ, PK and ML has received funding from the European Research Council (ERC) under the European Union’s Horizon 2020 research and innovation programme (grant agreement No. 851931). MDZ also acknowledges support from the Simons Foundation Grant \#888984 (Simons Collaboration on Global Categorical Symmetries). We thank the Simons Center for Geometry and Physics in Stony Brook (NY) for hospitality while completing this work.

\bibliographystyle{ytphys}
\bibliography{refs.bib}

\end{document}